\documentclass[iop]{emulateapj}

\usepackage{appendix,natbib}
\usepackage{amsmath}
\usepackage{subfigure}
\usepackage[backref,colorlinks=true,citecolor=blue,linkcolor=blue,urlcolor=blue]{hyperref}
\usepackage[all]{hypcap}

\usepackage{amsmath}
\usepackage{appendix}
\usepackage{url}

\newcommand{\xiion}{\ensuremath{\xi_{\mathrm{ion}}}}
\newcommand{\halpha}{H\ensuremath{\alpha}}
\newcommand{\hbeta}{H\ensuremath{\beta}}

\def\o32{{O$_{\rm 32}$}}
\def\msun{${\rm M_\odot}$}

\slugcomment{Accepted for publication in ApJ}

\begin{document}

\title{The {MOSDEF} Survey: Direct Observational Constraints on the Ionizing Photon Production Efficiency, {\xiion}, at $\lowercase{z}\sim 2$}

\author{\sc Irene Shivaei\altaffilmark{1,2}, Naveen A. Reddy\altaffilmark{1}, Brian Siana\altaffilmark{1}, Alice E. Shapley\altaffilmark{3}, Mariska Kriek\altaffilmark{4}, Bahram Mobasher\altaffilmark{1}, William R. Freeman\altaffilmark{1}, Ryan L. Sanders\altaffilmark{3}, Alison L. Coil\altaffilmark{5}, Sedona H. Price\altaffilmark{4,6}, Tara Fetherolf\altaffilmark{1}, Mojegan Azadi\altaffilmark{5}, Gene Leung\altaffilmark{5}, Tom Zick\altaffilmark{4}}

\altaffiltext{~}{\href{mailto:ishivaei@email.arizona.edu}{ishivaei@email.arizona.edu}}
\altaffiltext{1}{Department of Physics \& Astronomy, University of California, Riverside, CA 92521, USA}
\altaffiltext{2}{Steward Observatory, University of Arizona, Tucson, AZ 85721, USA}
\altaffiltext{3}{Department of Physics \& Astronomy, University of California, Los Angeles, CA 90095, USA}
\altaffiltext{4}{Astronomy Department, University of California, Berkeley, CA 94720, USA}
\altaffiltext{5}{Center for Astrophysics and Space Sciences, University of California, San Diego, La Jolla, CA 92093, USA}
\altaffiltext{6}{Max-Planck-Institut f${\rm \ddot u}$r extraterrestrische Physik, Giessenbachstr. 1, D-85737 Garching, Germany}

\begin{abstract}

We combine {\halpha} and {\hbeta} spectroscopic measurements and UV photometry for a  sample of 673 galaxies from the MOSDEF survey to constrain hydrogen-ionizing photon production efficiencies ({\xiion}) at $z=1.4-2.6$. We find $\langle\log(\xi_{\rm{ion}}/[{\rm{s^{-1}/erg\,s^{-1}\,Hz^{-1}}}])\rangle=25.06~(25.34)$, assuming the Calzetti (SMC) curve for the UV dust correction and a scatter of 0.28\,dex in {\xiion} distribution. 
After accounting for observational uncertainties and variations in dust attenuation, we conclude that the remaining scatter in {\xiion} is likely dominated by galaxy-to-galaxy variations in stellar populations, including the slope and upper-mass cutoff of the initial mass function, stellar metallicity, star-formation burstiness, and stellar evolution (e.g., single/binary star evolution).
Moreover, {\xiion} is elevated in galaxies with high ionization states (high [O{\sc iii}]/[O{\sc ii}]) and low oxygen abundances (low [N{\sc ii}]/{\halpha} and high [O{\sc iii}]/{\hbeta}) in the ionized ISM. However, {\xiion} does not correlate with the offset from the $z\sim0$ star-forming locus in the BPT diagram, suggesting no change in the hardness of ionizing radiation accompanying the offset from the $z\sim0$ sequence. We also find that galaxies with blue UV spectral slopes ($\langle\beta\rangle=-2.1$) have elevated {\xiion} by a factor of $\sim2$ relative to the average {\xiion} of the sample ($\langle\beta\rangle=-1.4$). If these blue galaxies are similar to those at $z>6$, our results suggest that a lower Lyman continuum escape fraction is required for galaxies to maintain reionization, compared to the canonical {\xiion} predictions from stellar population models.  Furthermore, we demonstrate that even with robustly dust-corrected {\halpha}, the UV dust attenuation can cause on average a $\sim0.3$\,dex systematic uncertainty in {\xiion} calculations. 

\end{abstract}


\maketitle

\section{Introduction}

Star-forming systems dominate the galaxy population at $z\gtrsim 1.5$, both in number and mass density \citep{madau14}. The light emitted from these galaxies is dominated by emission from massive stars, mainly in UV and re-processed IR emission.
Therefore, our understanding of galaxy evolution is intimately tied to the physics of massive stars and the surrounding gas that is ionized by their radiation.

An important parameter that is determined by the population of massive stars is the galaxy hydrogen-ionizing photon production rate, which impacts the derivation of physical properties from nebular emission lines, such as star-formation rates (SFRs) and gas-phase metallicities. The production rate of Lyman-continuum (LyC) ionizing photons per unit non-ionizing UV continuum luminosity, the latter usually measured at 1500\,\AA, is called the ionizing photon production efficiency, {\xiion} \citep[e.g.,][]{robertson13}.
{\xiion} depends on the ionizing photon production rate of massive stars, the spectral shape of the UV emission from hot massive stars, and the relative number of such stars to the number of less massive stars that do not produce appreciable ionizing radiation.
The fraction of massive ionizing stars is regulated by the initial mass function (IMF), star formation history, and age of the galaxy. 
Additionally, the ionizing photon production rate per unit stellar mass is dependent on the metallicity, binarity, and rotation of massive stars \citep[e.g.,][]{stanway16,choi17}. Hence, determining {\xiion} helps us to place constraints on the massive stellar populations in high-redshift galaxies and improve our knowledge of galaxy evolution and stellar population synthesis (SPS) models. Finally, {\xiion} is an important input in models of cosmic reionization and plays an important role in determining whether star-forming galaxies are capable of reionizing the universe \citep[e.g.,][]{bouwens15b,robertson15}.

In studies of high-redshift galaxies, {\xiion} is often {\em predicted} either directly from SPS models (for a given set of assumptions regarding the stellar properties of high-redshift galaxies; \citealt{madau99}) or by using the observed UV continuum slope $\beta$ as an indicator of {\xiion}, based on the $\beta$-{\xiion} relations that are derived from SPS models considering a variety of metallicities, ages, and dust content \citep[e.g.,][]{robertson13,bouwens15b,duncan15}.
The ``canonical'' {\xiion} value ($\log(\xi_{\mathrm{ion}}/{\rm erg\,Hz^{-1}})\sim 25.2-25.3$; \citealt{robertson13}) is derived from these indirect estimates.
A direct way of {\em measuring} {\xiion} is by modeling UV spectra and determining {\xiion} either from the UV metal lines \citep{stark15,stark17} or by evaluating the ratio of ionizing to non-ionizing fluxes ($f_{900}/f_{1500}$, \citealt{reddy16b}). 
The main advantage of these methods is that they are less sensitive to uncertainties in dust attenuation corrections. However, they require deep rest-UV spectroscopy, and the current results are limited to either individual extreme objects at high redshifts \citep{stark17} or the composite spectrum of a large number of galaxies \citep{reddy16b}. In both cases, statistical information on the galaxy-to-galaxy variations of {\xiion} in large samples is missed. 

Another method of measuring {\xiion} that enables us to constrain its value for individual galaxies is by using nebular recombination lines, e.g. {\halpha}. 
In an ionization-bounded nebula, the rate of recombinations balances the rate of photons with energies at or above 13.6\,eV that are either emitted from the star or produced during recombination to the hydrogen ground level.
There are a number of studies that have used Balmer emission lines to estimate {\xiion} at high redshifts. For example, \citet{bouwens16b} estimated {\xiion} using indirect {\halpha} measurements at $z\sim 4-5$ based on {\em Spitzer}/IRAC photometry.
\citet{nakajima16} derived {\xiion} values from Keck/MOSFIRE spectra of {\hbeta} for a sample of 15 galaxies at $z=3.1-3.7$ that included a few candidate LyC leakers identified on the basis of ground-based imaging.
The recent study of \citet{matthee17} used {\halpha} for a large sample of narrow-band selected {\halpha} and Lyman-$\alpha$ emitters to constrain {\xiion} at $z=2.2$. 
Despite these advances, almost all of these studies did not have access to additional Balmer lines to correct the nebular lines for dust obscuration \citep{bouwens16b,nakajima16,matthee17}. Additionally, previous studies have either relied on small sample sizes \citep{nakajima16}, or did not have spectroscopic measurements of the nebular emission lines \citep{bouwens16b,matthee17}. A detailed and thorough study of {\xiion} requires large and representative samples with spectroscopic {\halpha} and {\hbeta} observations to directly measure nebular line luminosity and properly correct for dust.

To that end, we use data from the MOSFIRE Deep Evolution Field (MOSDEF) survey \citep{kriek15}, for which near-IR spectra are available for a representative sample of 773 galaxies at $1.37\leq z\leq 2.61$ with coverage of {\halpha} and {\hbeta}. 
Access to both {\halpha} and {\hbeta} spectroscopic measurements enables us to measure robust, dust-corrected {\halpha} luminosities \citep{kennicutt09,hao11,reddy15,shivaei15b,shivaei16} and infer ionizing photon production rates.
Moreover, the wealth of rest-optical emission lines that are accessible through MOSDEF (e.g., [O{\sc ii}], [O{\sc iii}], [N{\sc ii}]) helps us to study trends of {\xiion} with other commonly measured properties of galaxies, such as gas-phase metallicity and ionization parameter \citep{sanders15,sanders16a,shapley15}.
Our sample is further accompanied by the rich 3D-{\em HST} photometric dataset \citep{skelton14}, which aids in modeling stellar population parameters (e.g., stellar reddening and stellar mass).
The main goals of this study are (a) to provide an observationally constrained canonical {\xiion} value for typical star-forming galaxies at $z\sim 2$, (b) to explore the origin of the observed scatter in the {\xiion} distribution of galaxies, and (c) to investigate the variations of {\xiion} with galaxy global properties, ISM characteristics, and redshift.

The layout of this paper is as follows. In Section~\ref{sec:data}, we introduce the MOSDEF survey and our methods for measuring emission line fluxes and modeling the stellar populations. In Section~\ref{sec:xi_meas}, we describe the details of {\xiion} measurements for individual galaxies and average values from composite spectra. 
In Section~\ref{sec:scatter}, we present the average {\xiion} of our sample, and discuss potential sources of the large scatter in {\xiion}, as well as the uncertainties associated with dust attenuation corrections.
We show in Section~\ref{sec:avg} the relationship between {\xiion} and UV continuum slope, UV magnitude, and line ratios, which are tracers of gas-phase metallicity and the hardness and intensity of the ionization field. We compare our measurements with those in the literature in Section~\ref{sec:z}. Finally, our results are summarized in Section~\ref{sec:summary}.
Throughout this paper, line wavelengths are presented in vacuum, and magnitudes are given in the AB system \citep{oke83}. We assume a \citet{chabrier03} IMF. A cosmology with $H_0 = 70\,{\rm km s^{-1} Mpc^{-1}, \Omega_{\Lambda} = 0.7}$, and ${\rm \Omega_m = 0.3}$ is adopted.

\section{Data and Measurements}
\label{sec:data}

\subsection{The MOSDEF Survey}
\label{mosdef}
The recently completed MOSDEF survey was a four and a half year near-IR spectroscopic survey that used the Keck/MOSFIRE spectrograph \citep{mclean12} to study the properties of $\sim 1500$ galaxies and active galactic nuclei (AGNs) at $1.4\leq z\leq 3.8$ in the five CANDELS fields: AEGIS, COSMOS, GOODS-N, GOODS-S, and UDS \citep{grogin11,koekemoer11}. The fields were also covered by the {\em HST}/WFC3 grism observations of the 3D-{\em HST} survey \citep{skelton14,momcheva16}. The MOSDEF parent sample was selected down to $H=24.0$, 24.5, and 25.0 at $z=1.37-1.70$, $2.09-2.61$, and $2.95-3.80$, respectively. 
For the details of the survey, observing strategy, and data reduction, we refer readers to \citet{kriek15}.

\subsection{Line Luminosities and Nebular Dust Correction}
\label{line_lum}
Emission line fluxes are estimated by fitting Gaussian functions to the 1D spectra, where the uncertainties are derived by perturbing the spectrum of each object according to its error spectrum and measuring the line fluxes in the perturbed spectra \citep{kriek15,reddy15}. Slit loss corrections are applied by normalizing the spectrum of a star in each observing mask to match the 3D-{\em HST} total photometric flux \citep{skelton14}. 
Additionally, we use {\em HST} images of our resolved galaxy targets to estimate and correct for differential flux loss relative to that of the star described above \citep{kriek15}. 
There are 767 galaxies (excluding AGNs on the basis of X-ray emission, IRAC colors, and rest-optical [N{\sc II}]/{\halpha} line ratio $>0.5$) with {\halpha} and {\hbeta} spectra. In 698 of these (91\%), at least one of the two lines is detected at 3\,$\sigma$ (673 have {\halpha} detection), and in 451 (59\%) both lines are detected. In this work, we use the 673 galaxies with {\halpha} detections.

We correct the H$\alpha$ and H$\beta$ fluxes for underlying Balmer absorption, determined from the best-fit SED models (Section~\ref{sec:sed}). 
To correct the nebular line luminosities for dust attenuation, we calculate the Balmer decrement ({\halpha}/{\hbeta}) and assume the \citet{cardelli89} Galactic extinction curve \citep{reddy15,shivaei15b,shivaei16}. 

\subsection{SED Fitting and Stellar Dust Correction}
\label{sec:sed}
We fit the rest-frame UV to near-IR photometry from 3D-{\em HST} \citep{skelton14,momcheva16}, corrected for emission line contamination according to the MOSDEF spectra. For this analysis, we use \citet[][hereafter BC03]{bc03} models, employing a minimum $\chi^2$ method to find the best-fit models and derive stellar masses and stellar color-excesses \citep[see][]{reddy15}.
For the SED models, we assume a \citet{chabrier03} IMF, an exponentially rising star formation history \citep[which has been shown to best reproduce the observed SFRs at $z\sim 2$,][]{reddy12b}, and two combinations of metallicities and attenuation curves: a solar metallicity with the \citet{calzetti00} curve as is commonly assumed in high-redshift studies, and a subsolar metallicity (0.2\,$Z_{\odot}$) with an SMC curve \citep{gordon03}, motivated by recent studies suggesting that a steeper attenuation curve with a subsolar metallicity ($\sim 0.14-0.20\,Z_{\odot}$) is more applicable to galaxies at high redshifts \citep{capak15,bouwens16c,reddy17}.
We measure the stellar $E(B-V)$ and UV luminosity ($L_{\nu}$) at 1500\,\AA~from the best-fit SED models. The UV slope ($\beta$) is also measured from the best-fit SED by fitting a power-law function ($f_{\lambda}\propto \lambda^{\alpha}$) at rest-frame wavelengths $1268-2580$\,\AA~(using the continuum windows introduced in \citealt{calzetti94}). The galaxies in our sample have at least three photometric bands in the aforementioned wavelength window, with an average of eight bands.

\section{Inferring \xiion}
\label{sec:xi_meas}
The LyC production efficiency ({\xiion}) is defined as the ratio of the production rate of ionizing photons ($N({\rm H}^0)$) in units of ${\rm s^{-1}}$ to the UV continuum luminosity density ($L_{\rm UV}$) in units of ${\rm erg\,s^{-1}\,Hz^{-1}}$:
\begin{equation}
\xiion = \frac{N({\rm H}^0)}{L_{\rm UV}}~ {\rm [s^{-1}/erg\,s^{-1}\,Hz^{-1}]}.
\label{eq:xi}
\end{equation}
In this work, $L_{\rm UV}$ is measured at 1500\,\AA~(Section~\ref{sec:sed}), and $N({\rm H}^0)$ is calculated as follows.
For an ionization-bounded nebula, where the rate of recombination balances the rate of ionizing photons, one can convert the {\halpha} luminosity to the production rate of LyC photons. This conversion mainly depends on the nebular conditions, including electron density and gas temperature \citep{spitzer78,osterbrock89}, and hence is not sensitive to the choice of stellar population model.
We use the relation of \citet{leithererheckman95} to convert our dust-corrected {\halpha} luminosity (Section~\ref{line_lum}) to $N({\rm H}^0)$:
\begin{equation}
N({\rm H^0})~ {\rm [s^{-1}]} = \frac{1}{1.36} \times 10^{12} L({\rm \halpha})~ {\rm[erg\,s^{-1}]}.
\label{eq:n0}
\end{equation}

An important assumption in calculating the LyC photon production rate based on the observed {\halpha} luminosity is the amount of dust attenuation of ionizing photons internal to H{\sc ii} regions. Using a sample of seven local galaxies, \citet{inoue02} concluded that the dust optical depth of LyC photons in H{\sc ii} regions is almost equal to the dust optical depth of the diffuse ISM\footnote{If we assume that the gas-to-dust ratio in our galaxies is similar to that of the local star-forming regions used in the \citet{inoue02} study (an assumption that is not necessarily correct), their result implies that $\sim 30\%$ of LyC photons in our galaxies (with $\langle E(B-V)\rangle \sim 0.2$) are depleted by dust attenuation internal to H{\sc ii} regions. To correct for this effect, one would have to increase the {\xiion} computed from the H$\alpha$ luminosity by $\sim 0.1$\,dex.}. On the other hand, at $z\sim 2$, \citet{reddy16a} used rest-UV spectroscopy and showed that the neutral hydrogen column densities are high enough where ionizing photons have a higher probability of ionizing hydrogen before they are absorbed by dust grains. Additionally, the overall agreement between {\halpha} and UV SFRs that has been shown in many previous studies \citep[e.g.,][]{erb06c,shivaei15a,shivaei15b} indicates that photoelectric absorption dominates the depletion of ionizing photons. Therefore, in this analysis, we assume negligible dust attenuation of ionizing photons internal to H{\sc ii} regions. 

Another assumption in the {\halpha}-derived production rate of ionizing photons is the escape fraction of LyC photons ($f_{\rm esc}$). A nonzero $f_{\rm esc}$ implies a larger intrinsic {\xiion} by a factor of $\frac{1}{(1-f_{\rm esc})}$ with respect to the measured {\xiion} based on {\halpha} luminosity. We assume the relation between the covering fraction of neutral hydrogen ($f_{\rm cov}$) and $E(B-V)$ from \citet{reddy16b} to estimate $f_{\rm esc}$ (i.e., $1 - f_{\rm cov}$) from our SED-inferred $E(B-V)$, and correct our measured {\xiion} values.
For the nine galaxies in our entire sample with $E(B-V)=0$ (four, if we use an SMC curve with 0.2 solar metallicity), we set $f_{\rm esc}=0.3$, which is the maximum value of the sample, to avoid an unrealistic escape fraction of 100\%. 
With the assumption of the Calzetti (SMC) attenuation curve for the SED models, the average $f_{\rm esc}$ for the full sample is 0.04 (0.09). Ignoring the mentioned $f_{\rm esc}$ correction (i.e., setting $f_{\rm esc}=0$) would not significantly alter the trends we find in this analysis.

Important sources of uncertainty in our inferences of {\xiion} are the dust corrections applied to the unobscured {\halpha} and UV luminosities. There are a number of studies in the literature that have measured the UV and optical attenuation of stellar and nebular emission at high redshifts \citep[][among many others]{noll09,wild11,reddy12b,reddy15,shivaei15a,debarros16,shivaei16}.  Corrections for the nebular attenuation commonly assume the Balmer decrement (when measured) and the \citet{cardelli89} Galactic extinction curve \citep{calzetti00,steidel14,reddy15,shivaei16}. The dust-corrected {\halpha} luminosity increases on average by $\sim 0.06$\,dex if we use the \citet{calzetti00} curve instead of the \citet{cardelli89} curve. Estimating the stellar UV dust attenuation is more uncertain. As mentioned in Section~\ref{sec:sed}, we used $E(B-V)$ from the best-fit SED models with the assumption of the Calzetti$+$solar metallicity or the SMC$+$subsolar metallicity. The effect of the UV dust correction on {\xiion} is discussed in Section~\ref{sec:dustcurve}.

\subsection{Composite Measurements}
\label{sec:stack}

In order to include {\hbeta}-undetected galaxies in our analysis, we use a stacking technique to derive average {\xiion} values in bins of galaxy parameters as follows.\footnote{Source code available at \url{https://github.com/IreneShivaei/specline}}
Individual spectra are shifted to the rest-frame with a wavelength grid where $\delta\lambda = 0.5$\,\AA. 
Parts of the individual spectra where the ratio of the error spectra to the median error is greater than 3 are removed, as they are dominated by sky lines. 
The composite spectrum is created by taking the mean of the luminosities of individual spectra contributing to the stack in each wavelength bin.
The composite error spectrum at each wavelength is estimated as the uncertainty on the weighted mean (i.e., $\frac{1}{\sqrt{\sum\limits_{i} \sigma_i^2}}$, where $\sigma_i^2$ is the variance of the $i$th individual spectrum).

We fit the composite {\halpha}, {\hbeta}, [O{\sc iii}], and [N{\sc ii}] lines with single Gaussian profiles and the [O{\sc ii}] line with a double Gaussian profile. We estimate the composite line errors by perturbing the stacked spectrum within the composite error spectrum 1000 times. The {\halpha} and {\hbeta} composite lines are corrected for Balmer absorption using the mean absorption of the individual galaxies contributing to the stack, inversely weighted by individual {\halpha} and {\hbeta} flux uncertainties, respectively.

To derive Balmer decrements, we initially normalize individual spectra to their {\halpha} luminosities, then create the composite spectrum from the {\halpha}-normalized spectra, and take the inverse value of the normalized composite {\hbeta} line as the mean $\langle \frac{\rm H\alpha}{\rm H\beta}\rangle$ ratio. In a similar manner, we create the UV-normalized composite spectra and measure the {\halpha} luminosities to derive the observed (uncorrected for dust) $\langle \frac{\rm H\alpha}{\rm UV}\rangle$ ratio, which is required in Equations~\ref{eq:xi} and \ref{eq:n0}. 

In Section~\ref{sec:gas}, we measure $\langle \frac{\rm [O{\normalfont\textsc{iii}}]}{\rm H\beta}\rangle$ and $\langle \frac{\rm [O{\normalfont\textsc{iii}}]}{\rm [O{\normalfont\textsc{ii}}]}\rangle$ from the composite spectra. The purpose of this calculation is to include objects that are undetected in {\hbeta} and [O{\sc ii}]. We create the composite spectra in a similar manner to that described above, by initially normalizing individual spectra to the [O{\sc iii}] luminosity, and then taking the average luminosities in each wavelength bin. The dust attenuation term for $\langle \frac{\rm [O{\normalfont\textsc{iii}}]}{\rm [O{\normalfont\textsc{ii}}]}\rangle$ is measured from the composite spectrum of the same set of objects, but this time normalized to {\halpha} (i.e., the $\langle \frac{\rm H\alpha}{\rm H\beta}\rangle$ ratio explained above).
We test the validity of our composite line measurements by stacking only the galaxies that are detected in the relevant lines and comparing the composite line ratio with the average of the individual line ratios. We find that the two agree well within the uncertainties.

\begin{figure*}[tbp]
	\centering
		\includegraphics[width=.75\textwidth,trim={.2cm 0 0 0},clip]{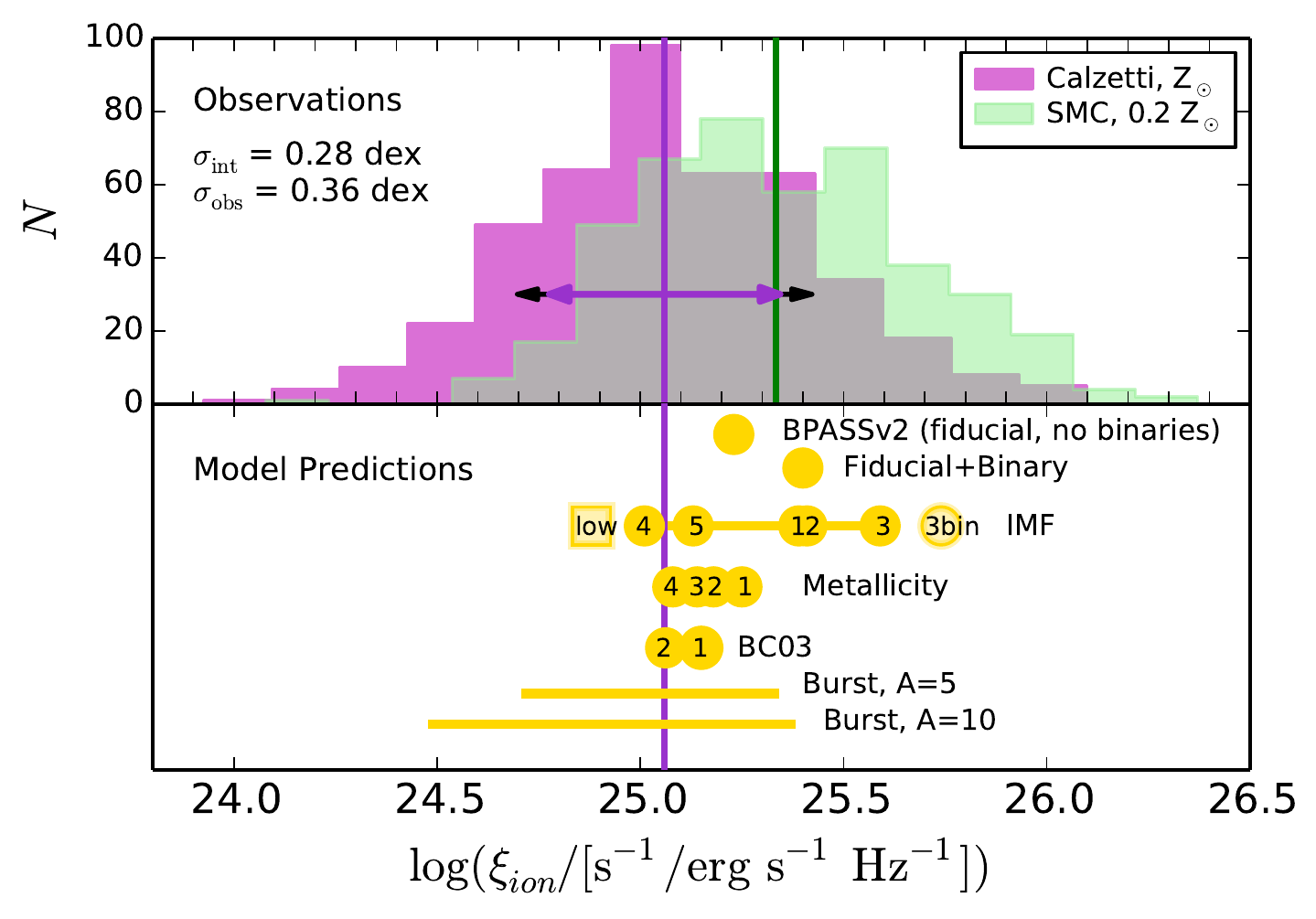}
		\caption{{\em Top:} distribution of {\xiion} for our entire sample of $1.4\leq z\leq 2.6$ galaxies, assuming the Calzetti attenuation curve for the UV dust correction in violet and an SMC curve in green. The observed and intrinsic (i.e., measurement-subtracted) scatters are 0.36 and 0.28\,dex for the Calzetti distribution and are shown with black and violet arrows, respectively. The vertical lines show the averages of the sample. Assuming an SMC curve systematically increases {\xiion} by $\sim 0.3$\,dex, indicating the sensitivity of {\xiion} measurements to the assumed UV attenuation curve.
		{\em Bottom:} {\xiion} predictions from the stellar population models of BPASSv2 and BC03. Each row shows the variation of {\xiion} by changing the labeled quantity in the model, described in detail in Table~\ref{tab:models}. The numbers refer to the model assumptions reported in Table~\ref{tab:models}. The light yellow circle (labeled 3bin, $\log(\xi_{\rm{ion}}/[{\rm{s^{-1}/erg\,s^{-1}\,Hz^{-1}}}])=25.74$) is our highest {\xiion} prediction with the IMF model \#3 in Table~\ref{tab:models} including binary evolution. The light yellow square is the lowest {\xiion} prediction (24.88) that assumes the IMF model \#4 but with a higher metallicity of $Z=0.02$.
		}
		\label{fig:scatter}
\end{figure*}

\section{{\xiion} Distribution}
\label{sec:scatter}

Figure~\ref{fig:scatter} shows the distribution of {\xiion} for the MOSDEF galaxies at $z=1.37-2.61$ with {\halpha} and {\hbeta} lines detected at 3\,$\sigma$ (top panel). The average {\xiion} is $\log(\xi_{\rm{ion}}/[{\rm{s^{-1}/erg\,s^{-1}\,Hz^{-1}}}])=25.10~(25.33)$, assuming the Calzetti (SMC) curve for the UV dust correction (excluding the {\hbeta} undetected objects results in an average {\xiion} of 25.06 and 25.34, assuming the Calzetti and SMC curves, respectively; {\xiion}$=25.06$ and 25.34 are shown with vertical lines in Figure~\ref{fig:scatter}).
The average {\xiion} for the subsamples at $z=1.4-2.0$ and $z=2.0-2.6$ are 25.05 (25.29) and 25.13 (25.36), assuming the Calzetti (SMC) curve.

The observed scatter in the {\xiion} distribution (i.e., the standard deviation of the distribution) is 0.36\,dex (0.35\,dex) using the Calzetti (SMC) attenuation curve (shown with the black arrow in Figure~\ref{fig:scatter}). The average measurement uncertainty, which is dominated by the uncertainty in the {\halpha} and {\hbeta} luminosities, is 0.22\,dex. Subtracting the measurement uncertainty from the observed scatter in quadrature results in an ``intrinsic'' scatter of 0.28\,dex (0.26\,dex) for the Calzetti (SMC) curve (the 0.28\,dex scatter is shown with the purple arrow in Figure~\ref{fig:scatter}).
In the following two subsections, we investigate the effect on the {\xiion} distribution resulting from the UV dust attenuation curve assumption (Section~\ref{sec:dustcurve}) and variations in massive star populations predicted from models (Section~\ref{sec:sps}). 

\subsection{Dust Attenuation Correction Uncertainties}
\label{sec:dustcurve}

As shown in the top panel of Figure~\ref{fig:scatter}, using an SMC curve with subsolar metallicity systematically increases {\xiion} by $\sim 0.1-0.4$\,dex, with a $\sim 0.2$\,dex shift in the average values. The assumption of an SMC curve with subsolar metallicity is motivated by recent studies that suggest an SMC curve with a bluer intrinsic UV slope, $\beta_0$ (and hence, a higher attenuation at 1500\,\AA), is more appropriate for high-redshift galaxies \citep{reddy17}. 
The discrepancy between the results of the SMC and Calzetti assumptions increases with increasing mass and dustiness (See Figures~\ref{fig:beta} and \ref{fig:mass}). 
The average {\xiion} assuming the Calzetti curve and that assuming the SMC curve bracket the commonly used ``canonical'' {\xiion} of $10^{25.20}\,{\rm{s^{-1}/erg\,s^{-1}\,Hz^{-1}}}$, which is predicted from SPS models \citep{robertson13}. A {\xiion} that is different from the canonical value has important implications for the studies of cosmic reionization. For example, a higher (lower) {\xiion} results in a lower (higher) LyC photon escape fraction required for galaxies to maintain cosmic reionization \citep[we will return to this point in Section~\ref{sec:uv}, also see][]{robertson13,robertson15,bouwens16b,stanway16}.

Assuming a single attenuation curve for all types of galaxies with such a wide range of masses ($M_*\sim 10^{9-11}$\,{\msun}) and dustiness (UV slope $\beta\sim -2.4-1.2$) is probably not realistic. If the shape of the dust attenuation curve varies with galaxy properties, such as specific SFR or dust reddening \citep[as has been shown in other studies, such as][]{noll09,kriek13,reddy15,salmon16}, this variation will affect the trends we find between {\xiion} and galaxy parameters. 
For example, it has been suggested that for young galaxies, a curve that is steeper than the Calzetti curve provides a better agreement with IR observations \citep{reddy06a,siana09,reddy10,shivaei15a}. The more recent study by \citet{reddy17} suggested an SMC curve for massive galaxies and an attenuation curve steeper than SMC for young and low-mass galaxies at $z\sim 2$. A steeper attenuation curve leads to lower UV dust attenuation and higher {\xiion}. For simplicity, in this study we use a single attenuation curve for the entire sample and present results based on both the Calzetti and SMC curves throughout the analysis. Future UV and IR observations are needed to shed more light on this issue.

\begin{figure}[tbp]
	\centering
		\includegraphics[width=.45\textwidth,trim={.2cm 0 0 0},clip]{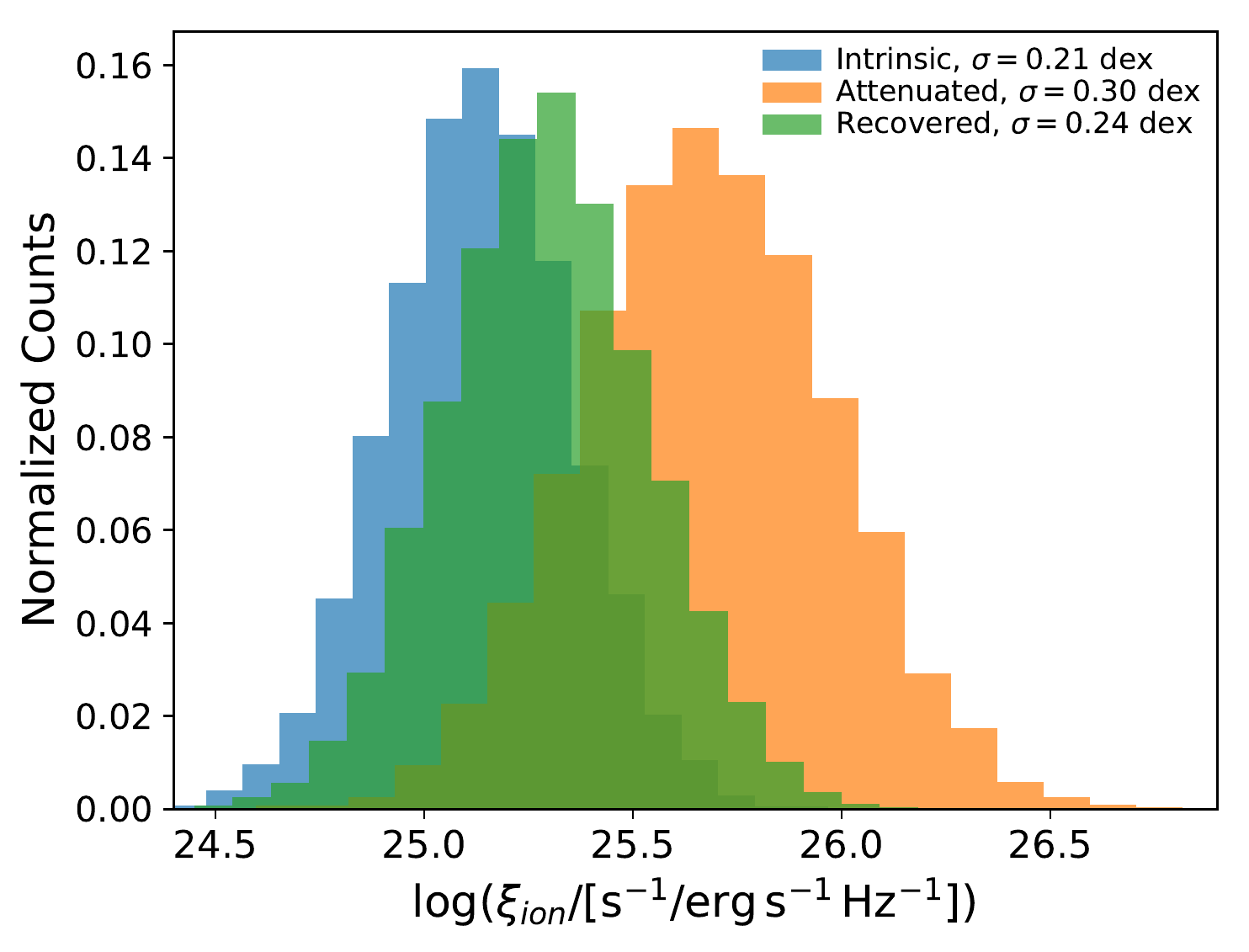}
		\caption{Simulation of a {\xiion} distribution with an intrinsic scatter of 0.21\,dex. The simulated intrinsic distribution is shown in blue. To test the effect of the galaxy-to-galaxy variations of the dust attenuation curve on the {\xiion} scatter, we attenuated each object with a different attenuation curve (orange) and corrected the dust-obscured luminosities this time by assuming a single attenuation curve (green). The recovered scatter is larger than the intrinsic one, which introduces a $\sim 0.1$\,dex uncertainty in the measured scatter of the {\xiion} distribution.
		}
		\label{fig:dustsim}
\end{figure}

\begin{figure}[tbp]
	\subfigure{
	\centering
		\includegraphics[width=.49\textwidth,trim={.2cm 0 0 0},clip]{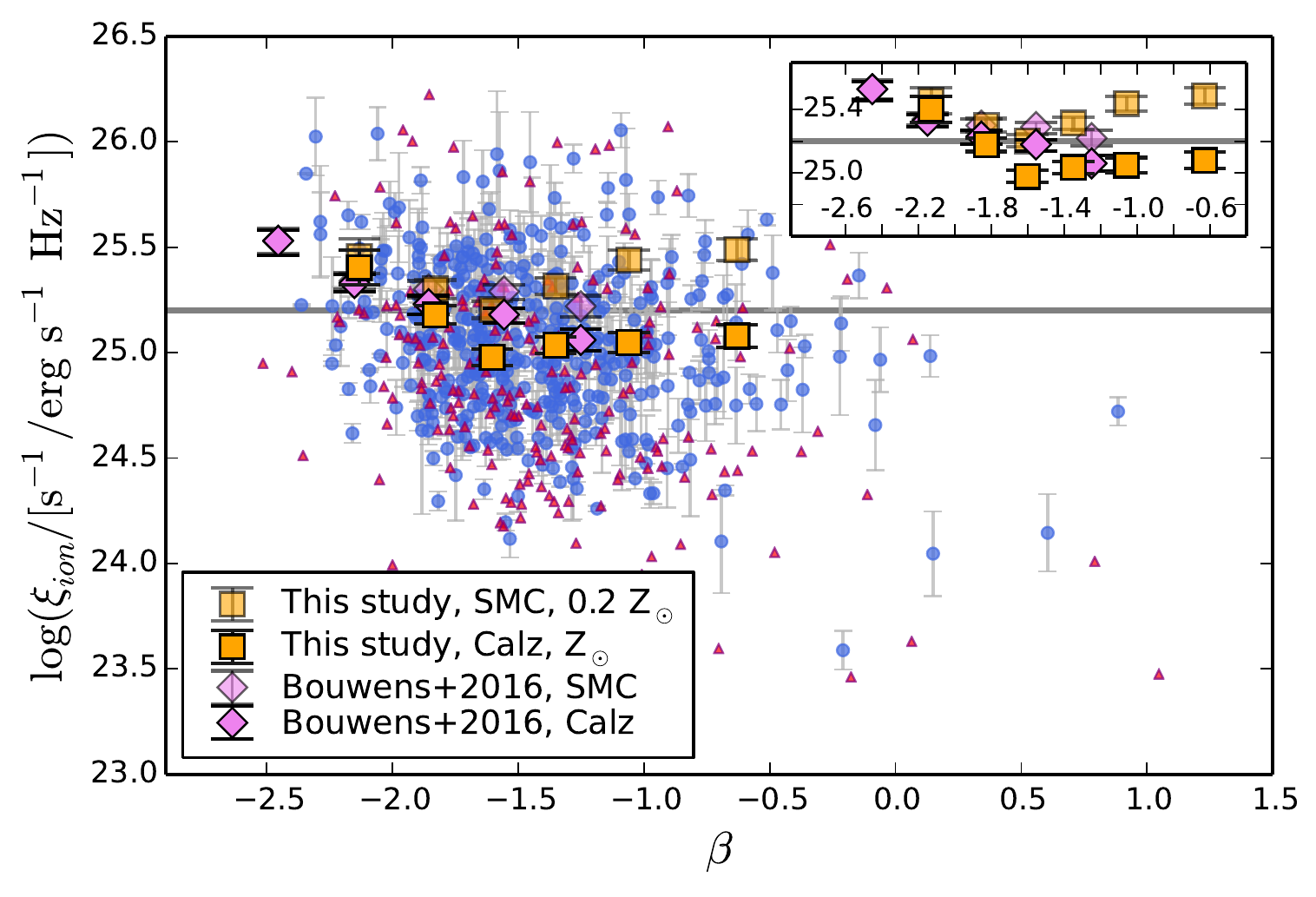}}
	\\
	\subfigure{
	\centering
		\includegraphics[width=.49\textwidth,trim={.2cm 0 0 0},clip]{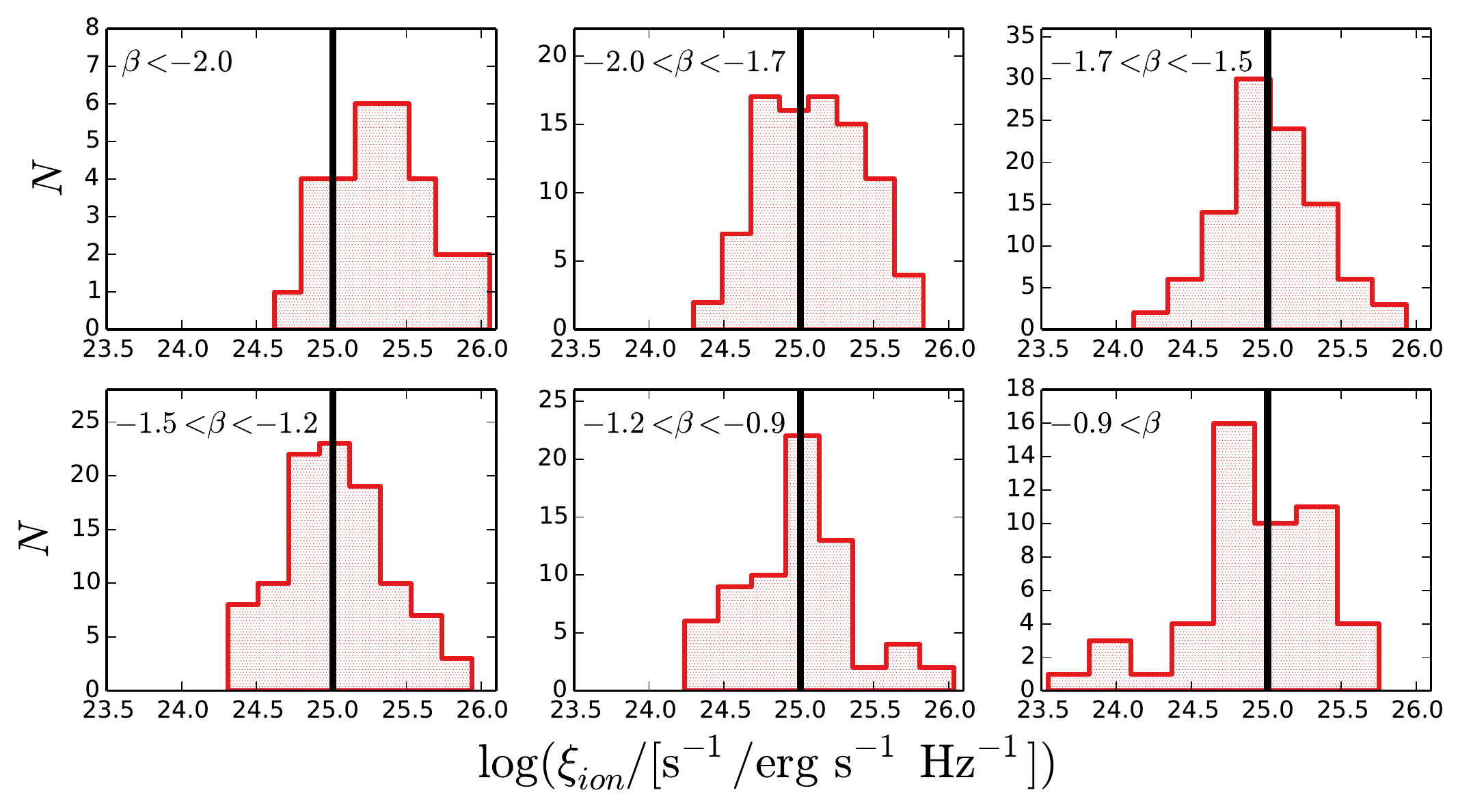}}
		\caption{ ({\em Top:}) Production efficiency of ionizing photons as a function of UV continuum slope, $\beta$. Small blue circles and red upward triangles show individual {\xiion} values for the {\halpha}-{\hbeta} detected sample, and $3\,\sigma$ lower limits for the {\hbeta}-undetected sample, respectively. 
		We use the Calzetti curve and a solar metallicity for the UV dust correction and the Cardelli curve for the {\halpha} dust correction (Equation~\ref{eq:xi}). Orange squares show the results of the composite spectra in bins of $\beta$ for the full sample (detected and undetected {\hbeta}). The dark orange stacks assume the Calzetti curve with a solar metallicity for the UV dust correction, and the light orange stacks assume an SMC curve with subsolar ($0.2\,Z_{\odot}$) metallicity. 
		Also shown is the result of the \citet{bouwens15b} study of $z=3.8-5.0$ galaxies in dark and light violet diamonds for the Calzetti and SMC curves with a solar metallicity ($\beta_0=-2.23$), respectively. As they did not have {\hbeta} measurements, \citet{bouwens15b} used stellar reddening and the same attenuation curve for nebular and stellar light.
		The horizontal line shows $\log(\xi_{\rm{ion}}/[{\rm{s^{-1}/erg\,s^{-1}\,Hz^{-1}}}])=25.2$ from \citet{robertson13}.
		({\em Bottom:}) Distribution of {\xiion} for the sample of {\halpha}-{\hbeta} detected galaxies in six bins of $\beta$ shown in the top plot. {\xiion} is measured using the Calzetti curve for the UV attenuation. The black vertical  line indicates the average {\xiion} of the full sample. It is clear that {\xiion} is elevated in the bluest galaxies.
		}
		\label{fig:beta}
\end{figure}

Other than the systematic change in {\xiion}, the galaxy-to-galaxy variation of dust attenuation curve is a source of uncertainty in the {\xiion} scatter. 
To estimate the uncertainty caused by the assumption of a single attenuation curve for all galaxies (the Calzetti or SMC for the stellar attenuation and the Cardelli for the nebular attenuation), we perform a simple test as follows. We simulate distributions of dust-corrected intrinsic {\halpha} and UV luminosities and calculate the distribution of {\xiion} with an intrinsic scatter of 0.21\,dex and an average value of $\log(\xi_{\rm{ion}}/[{\rm{s^{-1}/erg\,s^{-1}\,Hz^{-1}}}])=25.13$ (blue histogram in Figure~\ref{fig:dustsim}). 
The average {\halpha} and UV luminosities are matched to those of the dust-corrected {\halpha} and UV luminosity distributions in our sample. 
The intrinsic scatters of the {\halpha} and UV luminosity distributions assumed in the simulation (and hence, the intrinsic {\xiion} scatter of 0.21\,dex) are set such that later we recover the scatter in the {\em observed} {\xiion} distribution, uncorrected for dust (i.e., the {\xiion} inferred from the observed, uncorrected {\halpha} and UV luminosities), as follows.
Given the {\halpha} and UV luminosity of each galaxy, we predict the stellar and nebular $E(B-V)$. For this analysis, we fit the relations (and measure the scatters) of $L$({\halpha})$_{\rm corr}$ versus $E(B-V)_{\rm nebular}$ and $L$(UV)$_{\rm corr}$ versus $E(B-V)_{\rm stellar}$, based on the MOSDEF sample.
We attenuate each individual galaxy with a different attenuation curve ($\kappa_{\lambda}$), randomly picked from a uniform distribution of $\kappa_{\lambda}$ values.
The minimum and maximum $\kappa_{\lambda}$'s are chosen such that the range covers the commonly used attenuation curves in the literature. For the $L$({\halpha}) dust correction, we randomly pick a value between the $\kappa_{6565{\rm \AA}}$ of the \citet{reddy15} curve and that of the Calzetti curve. For the UV dust correction, the $\kappa_{1500{\rm \AA}}$ is randomly picked from a range between that of the \citet{reddy15} and the SMC curves. The result is shown as the orange histogram with a 0.30\,dex scatter in Figure~\ref{fig:dustsim}, which is similar to the scatter in the observed uncorrected {\xiion} distribution (0.28\,dex). In the final step, we recover the dust-corrected values, this time by assuming a single attenuation curve for all galaxies---the Calzetti curve for UV and the Cardelli curve for {\halpha}. The recovered scatter is 0.24\,dex, shown with the green histogram in Figure~\ref{fig:dustsim}. The recovered histogram also introduces a bias of $0.16$\,dex toward higher {\xiion}.

By subtracting, in quadrature, the intrinsic scatter (0.21\,dex) from the recovered scatter (0.24\,dex), we calculate an uncertainty of $\sim 0.1$\,dex associated with the uncertainties in dust corrections. The 0.1\,dex uncertainty is smaller than the 0.28\,dex scatter in {\xiion}. However, a caveat of the simple simulation presented here is that the dust attenuation curve (i.e., $\kappa_{\lambda}$) is not randomly distributed among galaxies. In fact, the slope of the attenuation curve is shown to correlate with the galaxy properties \citep[e.g.,][]{noll09,kriek13,reddy16a,salmon16}. A correlation between the attenuation curve steepness and the galaxy properties, such as mass, will result in a different (and potentially larger) broadening of the scatter in {\xiion}. We repeat our simulations for the case where the UV luminosity is corrected with an SMC curve at $E(B-V)_{\rm stellar} < 0.3$ and with the shallower Calzetti curve at $E(B-V)_{\rm stellar} > 0.3$, motivated by the results of \citet{salmon16}, and find a larger uncertainty of 0.2\,dex. A similar result is achieved if we assume an SMC curve for $M_* < 10^{10}$\,{\msun} and the Calzetti curve for $M_* > 10^{10}$\,{\msun}, motivated by the study of \citet{reddy16a}.
Even a larger uncertainty of 0.2\,dex cannot fully account for the scatter in {\xiion}.
In the next subsection, we discuss other possible sources of the scatter in {\xiion}, aside from the measurement uncertainties and the broadening uncertainty due to the galaxy-to-galaxy variations in the attenuation curve.

\subsection{Galaxy-to-galaxy Variations in Properties of Massive Star Populations}
\label{sec:sps}

Different properties of massive star populations, such as the shape of the IMF, starburstiness (i.e., bursts of star formation, as opposed to continuous star formation), metallicity, and binarity result in different ionizing photon production efficiencies. 
Here, we explore the predictions from the SPS models of BC03 and the BPASSv2 models \citep{eldridge16,stanway16} by adjusting their stellar population properties. 
To determine the predicted {\xiion} from an SPS model, we integrate $\lambda L_{\lambda}$ below 912\,\AA~and divide by $hc$, where $h$ is Planck's constant and $c$ is the speed of light, to convert the integrated value to the rate of ionizing photons, and divide the rate of ionizing photons by the UV luminosity density ($L_\nu$) at 1500\,\AA. The predicted values for a range of IMF parameters, metallicities, and the star-formation histories, and the effect of binaries are shown in Table~\ref{tab:models} and the bottom panel of Figure~\ref{fig:scatter}.

The commonly used BC03 model for a Salpeter IMF, a constant star formation history, and $Z=0.02$ ($1.5\,Z_{\odot}$, where $Z_{\odot}=0.0134$ based on the most recent solar abundances; \citealt{asplund09}) predicts $\log(\xi_{\rm{ion}}/[{\rm{s^{-1}/erg\,s^{-1}\,Hz^{-1}}}])=25.06$. The BPASSv2 model with the same assumptions produce a similar value of $\log(\xi_{\rm{ion}}/[{\rm{s^{-1}/erg\,s^{-1}\,Hz^{-1}}}])=25.08$.

Binaries are important sources of ionizing photons. In a binary system, the more massive star transfers mass and angular momentum to its less massive companion (an O- or B-type star), which increases the spin of the companion and leads to the rotational mixing of its layers and more efficient burning of hydrogen in its interior \citep[e.g.,][]{cantiello07,demink13}. Consequently, the main-sequence lifetime increases, and the star becomes hotter, which increases {\xiion} compared to that of a single star evolution \citep{stanway16}. Our fiducial BPASSv2 model without binary evolution, assuming a Salpeter IMF, a constant star formation history, and a subsolar metallicity ($0.15\,Z_{\odot}$) predicts $\log(\xi_{\rm{ion}}/[{\rm{s^{-1}/erg\,s^{-1}\,Hz^{-1}}}])=25.23$. Including binaries in the fiducial model leads to a {\xiion} higher by $\sim 50\%$ ($\log(\xi_{\rm{ion}}/[{\rm{s^{-1}/erg\,s^{-1}\,Hz^{-1}}}])=25.40$, Table~\ref{tab:models}).

Bursts of star formation result in the largest change of {\xiion}, from $\log(\xi_{\rm{ion}}/[{\rm{s^{-1}/erg\,s^{-1}\,Hz^{-1}}}])=24.7$ to 25.3 for a burst of 5\,\msun\,yr$^{-1}$ on top of a constant star formation history of 1\,\msun\,yr$^{-1}$. Due to different timescales of {\halpha} and UV light \citep{kennicutt12,shivaei15b}, at the onset of a burst, there are more ionizing photons produced at a given UV luminosity and hence a larger {\xiion}, while after the end of the burst, the ratio of ionizing to non-ionizing fluxes decreases.
A burst with a higher amplitude of 10\,\msun\,yr$^{-1}$ would cause a larger variation in {\xiion}, but such strong bursts are not likely for the mass range of our galaxies \citep[e.g.][]{hopkins14,dominguez15}. 

Variations in the stellar metallicity also change {\xiion}, with lower metallicity models producing higher {\xiion}. 
Low-metallicity stars have higher effective temperatures and harder ionizing spectra due to less metal blanketing, an effect through which high-energy photons are absorbed by metals in stellar atmospheres. 
Also, a larger mass of hydrogen and lower mass-loss rate due to radiation pressure-driven winds lead to longer main-sequence lifetimes of low-metallicity massive stars \citep{maeder94,vink01}.
Moreover, at low metallicity, the Hayashi limit, an almost vertical line on the HR diagram that corresponds to the largest radius that a star at a given mass can have in order to maintain hydrostatic equilibrium, moves to higher temperatures \citep{elias85}.
In our models, changing the stellar metallicity from $0.07\,Z_{\odot}$ to $1.5\,Z_{\odot}$ (based on the most recent solar abundances; \citealt{asplund09}), results in a $\sim 50\%$ change in ionizing photon production efficiency ({\xiion}$=25.25$ and 25.08, respectively).

Variations in the slope and upper-mass cutoff of the IMF also translate into significant changes in {\xiion}. We vary the slope of the high-mass end from $-2.00$ to $-2.70$, and consider both 100 and 300\,\msun~upper-mass cutoffs. The latter is motivated by observational evidence for the presence of stars with masses $> 150$\,\msun~in nearby star clusters \citep{crowther10,smith16}. These variations in the slope and upper-mass cutoff of the IMF cause a 0.58\,dex change in {\xiion} (from {\xiion}$= 25.01$ to 25.59; see Table~\ref{tab:models}).
An IMF with a shallower slope at the high-mass end and/or a higher upper-mass cutoff results in a larger fraction of hot and massive stars. The highest {\xiion} value in our assumed parameter space belongs to a binary model with the shallowest IMF slope ($\alpha = -2.0$), upper-mass cutoff of 300\,\msun, and a subsolar metallicity ($0.07\,Z_{\odot}$). The predicted {\xiion} for this model is $\log(\xi_{\rm{ion}}/[{\rm{s^{-1}/erg\,s^{-1}\,Hz^{-1}}}])=25.74$ and is shown with a light yellow circle in Figure~\ref{fig:scatter}.
The lowest {\xiion} value in our parameter space (excluding the starburst models, as they are highly dependent on the assumed burst amplitude and can reach to very low {\xiion} values) belongs to the model with the highest metallicity ($Z=0.02$), a steep IMF ($\alpha=-2.7$) with a 100\,{\msun} upper-mass cutoff, and no binary evolution. The lowest {\xiion} is $\log(\xi_{\rm{ion}}/[{\rm{s^{-1}/erg\,s^{-1}\,Hz^{-1}}}])=24.88$, shown with a light yellow square in Figure~\ref{fig:scatter}. 

In summary, in Section~\ref{sec:dustcurve} we show that the galaxy-to-galaxy variations of the dust attenuation curve cause a $\sim 0.1-0.2$\,dex scatter in {\xiion}. In this section, we conclude that the larger 0.28\,dex scatter in {\xiion} is partly affected by the variations in the stellar population of individual galaxies, such as the variations in the IMF, stellar metallicity, star formation history, and the binary/single evolutionary pathways of massive stars. However, the assumed dust attenuation curve for the UV stellar continuum correction causes a large {\em systematic} uncertainty of $\sim 0.2-0.3$\,dex, as shown with the green and violet histograms in Figure~\ref{fig:scatter}. Given the large scatter and 0.3\,dex systematic uncertainty in our measurements, as well as the large variations within the model predictions, it is not trivial to determine which model (or which dust attenuation curve) best describes the average properties of galaxies at $z\sim 2$ with the current data.

\def\arraystretch{1.2}
\capstartfalse   
\begin{deluxetable*}{cccccc}
\setlength{\tabcolsep}{0.5in} 
\tabletypesize{\footnotesize} 
\tablewidth{0pc}
\tablecaption{{\xiion} Predictions from SPS Models}
\tablehead{
\colhead{SPS Model} &
\colhead{IMF} &
\colhead{SFH} &
\colhead{Z} &
\colhead{Binary} &
\colhead{$\log(\xi_{\rm{ion}}/[{\rm{s^{-1}/erg\,s^{-1}\,Hz^{-1}}}])$}
}
\startdata
{BPASSv2} & {$\tablenotemark{i}\alpha= -2.35$, $M_*^{\rm max}=100$\,{\msun}} & {Constant over 300\,Myr} & {0.002} & {No} & {~~~25.23} \\
{(Fiducial, F)} & {} & {} & {\tiny (0.15\,$Z_{\odot}$)} & {} & {}\\
\hline
{(F)} & {(F)} & {(F)} & {(F)} & {Yes} & {~~~25.40}\\
\hline
{(F)} & {$\alpha=-2.35$, $M_*^{\rm max}=300$\,{\msun}} & {(F)} & {(F)} & {Yes} & {~~~25.59}\\
\hline
{(F)} & {$\left\{\begin{tabular}{@{\ }l@{}}
    (1) $\alpha=-2.35$, $M_*^{\rm max}=300$\,{\msun} \\ 
    (2) $\alpha=-2.00$, $M_*^{\rm max}=100$\,{\msun} \\ 
    (3) $\alpha=-2.00$, $M_*^{\rm max}=300$\,{\msun} \\ 
	(4) $\alpha=-2.70$, $M_*^{\rm max}=100$\,{\msun} \\ 
	(5) $\alpha=-2.70$, $M_*^{\rm max}=300$\,{\msun}
  \end{tabular}\right.$} & {(F)} & {(F)} & {(F)} & {$\left\{\begin{tabular}{@{\ }l@{}}
	25.39 \\
	25.41 \\
	25.59\tablenotemark{*} \\
	25.01\tablenotemark{*} \\
	25.13
  \end{tabular}\right.$
  }\\
\hline
{(F)} & {(F)} & {(F)} & {$\left\{\begin{tabular}{@{\ }l@{}}
	(1) 0.001 \\
	(2) 0.006 \\
	(3) 0.01 \\
	(4) 0.02
  \end{tabular}\right.$
} & {(F)} & {$\left\{\begin{tabular}{@{\ }l@{}}
	25.25 \\
	25.18 \\
	25.14 \\
	25.08
  \end{tabular}\right.$
}\\
\hline
{BC03} & {(F)} & {(F)} & {$\left\{\begin{tabular}{@{\ }l@{}}
	(1) 0.004 \\
	(2) 0.02
  \end{tabular}\right.$
} & {(F)} & {$\left\{\begin{tabular}{@{\ }l@{}}
	25.15 \\
	25.06
  \end{tabular}\right.$
}\\
\hline
{BC03} & {(F)} & {Burst $\left\{\begin{tabular}{@{\ }l@{}}
	(1) \tablenotemark{ii}$A$=5, $D$=100\,Myr \\
	(2) $A$=10, $D$=100\,Myr
  \end{tabular}\right.$
  } & {0.004} & {(F)} & {$\left\{\begin{tabular}{@{\ }l@{}}
	$[24.72-25.33]$ \\
	$[24.49-25.37]$
  \end{tabular}\right.$
  }
\enddata

\tablenotetext{}{From left to right, the columns indicate the SPS model, IMF, star formation history, metallicity, whether binary stars are included in the models, and the predicted {\xiion}. Entries with (F) mean the parameter is the same as the fiducial model (first row).}
\tablenotetext{i}{$\alpha$ is the IMF slope from 0.5\,{\msun} to $M_*^{\rm max}$, the upper-mass cutoff.}
\tablenotetext{ii}{$A$ and $D$ are the amplitude and the duration of the burst, respectively. The bursts are simulated on top of a constant star formation rate of 1\,{\msun}\,yr$^{-1}$. For example, $A=5$ means there is a burst with an amplitude of five times the underlying SFR.}
\tablenotetext{*}{The highest {\xiion} value in our parameter space is $\log(\xi_{\rm{ion}}/[{\rm{s^{-1}/erg\,s^{-1}\,Hz^{-1}}}])=25.74$ for the BPASSv2 model with binary evolution, an IMF slope of $-2.0$, upper-mass cutoff of 300\,\msun, and $Z=0.001$. The value is shown with a light yellow circle marker labeled 3bin in Figure~\ref{fig:scatter}.
The lowest {\xiion} value in our parameter space is $\log(\xi_{\rm{ion}}/[{\rm{s^{-1}/erg\,s^{-1}\,Hz^{-1}}}])=24.88$ for the BPASSv2 model without binary evolution, an IMF slope of $-2.7$, upper-mass cutoff of 100\,\msun, and $Z=0.02$. The value is shown with a light yellow square marker labeled low in Figure~\ref{fig:scatter}.}
\label{tab:models}
\end{deluxetable*}
\capstarttrue  

\begin{figure}[tbp]
	\centering
		\includegraphics[width=.49\textwidth,trim={.2cm 0 0 0},clip]{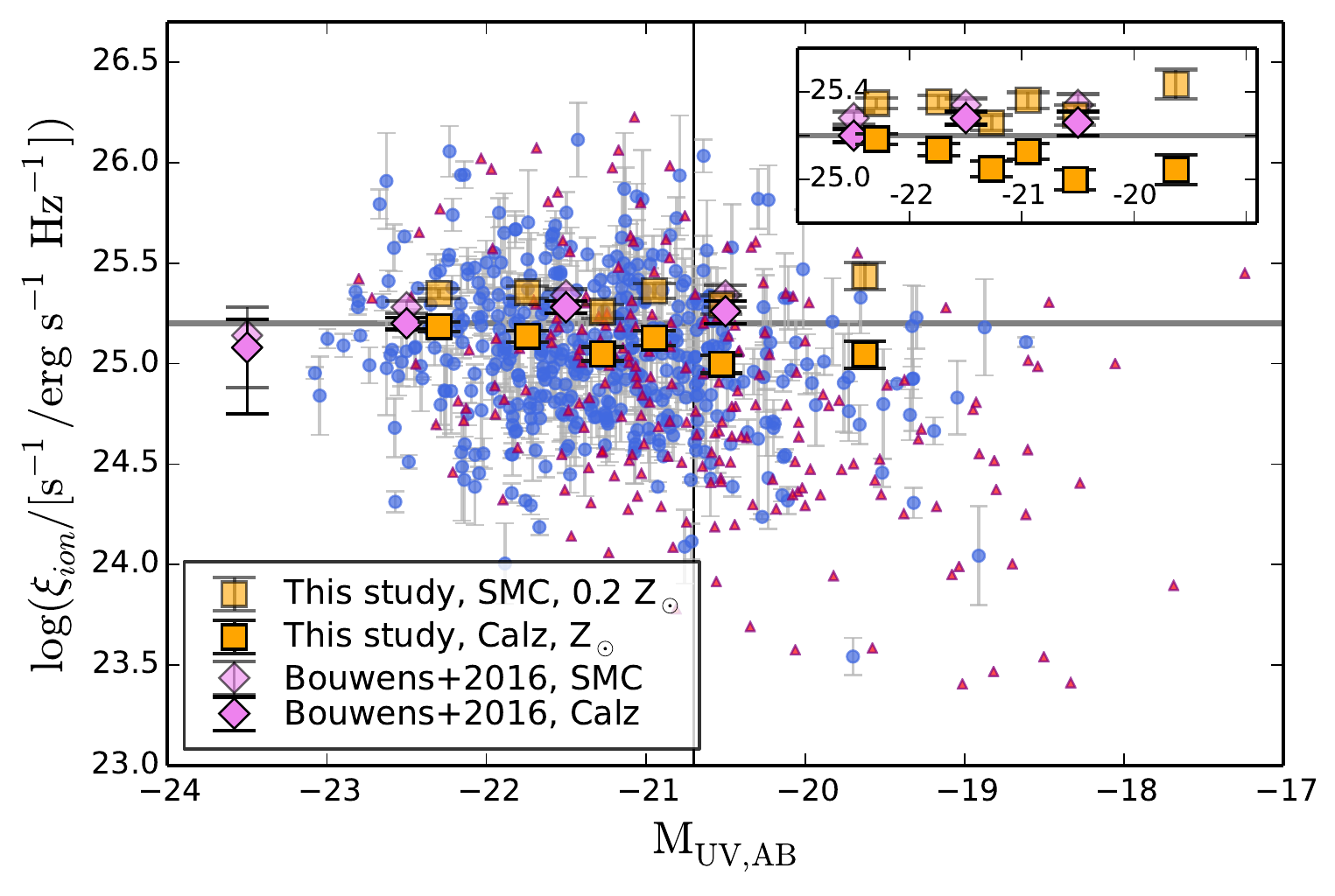}
		\caption{Dependence of {\xiion} on UV (1500\AA) absolute magnitude.
		The symbols are the same as in Figure~\ref{fig:beta}. The vertical line shows the characteristic UV magnitude at $1.9\leq z<2.7$ from \citet{reddy09}.
		}
		\label{fig:muv}
\end{figure}

\section{{\xiion} Evolution with Galaxy Properties}
\label{sec:avg}

In spite of the large scatter in {\xiion} discussed in Section~\ref{sec:scatter}, we can explore its average variation as a function of galaxy properties to learn more about the ionizing photon production efficiency in various environments. 
In this section, we evaluate the variations of {\xiion} as a function of stellar and gas properties in Sections~\ref{sec:uv} and \ref{sec:gas}, respectively. 

\subsection{{\xiion} and UV properties, implications for reionization}
\label{sec:uv}

At high redshift, the UV continuum slope ($\beta$) is often used as a proxy for the ionizing photon production efficiency \citep[e.g.,][]{robertson13,duncan15}, mainly because the rest-frame UV is shifted to longer optical wavelengths, which are easily accessible through ground-based observations across a large range of redshifts. 
Here, we have an independent measure of {\xiion} using nebular emission lines and can investigate the correlation between {\xiion} and the observed $\beta$, which is calculated by fitting a power law ($f_{\lambda}\propto \lambda^{\alpha}$) to the photometry at rest wavelengths $1268-2580$\,\AA. On average, our objects have eight photometric bands (at least three) in the rest-frame UV wavelength range.

In the top panel of Figure~\ref{fig:beta}, we show the relation between {\xiion} and $\beta$ for our sample (orange squares, $z=1.4-2.6$) and for the higher-redshift sample of \citet[][violet diamonds, $z=3.8-5.0$]{bouwens16b}.\footnote{The \citet{bouwens16b} study did not have {\hbeta} measurements, and they corrected their photometric-based {\halpha} luminosities using the stellar $E(B-V)$ and assuming a similar reddening toward nebular and stellar regions.}
Based on the {\xiion} values that are derived assuming the Calzetti curve for the UV dust correction (solid orange squares), {\xiion} is constant at $\beta \gtrsim -1.6$, and there is a significant increase of {\xiion} by $\sim 0.5$\,dex in the bluest bin with $\beta=-2.1$ ($\log(\xi_{\rm{ion}}/[{\rm{s^{-1}/erg\,s^{-1}\,Hz^{-1}}}])=25.41\pm 0.08$). This increase is consistent with the higher-redshift results of \citet{bouwens16b}, where they also infer an elevated {\xiion} in the galaxies with the bluest $\beta$. In the bottom panel of Figure~\ref{fig:beta}, we show the {\xiion} distribution in bins of $\beta$ with the vertical black line indicating the average of the full sample. Despite the large scatter in the individual measurements, it is evident that the {\xiion} distribution shifts to a lower mean value as $\beta$ becomes redder.

Studies of cosmic reionization struggle to show whether faint galaxies at $z>6$ have high enough escape fractions to maintain reionization \citep[e.g.,][]{vanzella10,siana15}. If the galaxies with the bluest $\beta$ ($\langle{\beta}\rangle=-2.1$) in our sample are similar to faint galaxies at $z>6$, their elevated {\xiion} of $\log(\xi_{\rm{ion}}/[{\rm{s^{-1}/erg\,s^{-1}\,Hz^{-1}}}])=25.41\pm 0.08$---compared to the ``canonical'' {\xiion} value of 25.20 that is commonly assumed in reionization models \citep{robertson13}---implies that a lower escape fraction of ionizing photons is needed for galaxies to reionize the universe \citep[see][]{robertson13,robertson15,bouwens16b,stanway16}. Even without the assumption of the $E(B-V)$-based $f_{\rm esc}$ (see Section~\ref{sec:xi_meas}), {\xiion} in the bluest bin is $25.32\pm 0.08$ for the Calzetti curve and $25.37 \pm 0.08$ for an SMC curve, which is $\sim 0.15$\,dex larger than the canonical \citet{robertson13} value.
The trends of {\xiion} with beta will be enhanced if the dust attenuation curve varies with galaxy properties in such a way that the curve steepens with decreasing mass and reddening \citep[as suggested by][]{reddy16a,salmon16}.

Using the constraints on the Thomson scattering and ionized hydrogen filling factor at $z=6-9$ inferred from CMB observations, \citet{bouwens15b} estimated the evolution of the cosmic ionizing emissivity. The cosmic ionizing emissivity, or the number density of ionizing photons per second capable of reionizing hydrogen in the IGM, $\dot{N}_{\rm ion}$, is a product of three terms: the total UV luminosity density ($\rho_{\rm UV}$), the ionizing photon production efficiency ({\xiion}), and the fraction of the ionizing photons that escape into the IGM ($f_{\rm esc}$). \citet{bouwens15b} compared the observationally constrained evolution of $\dot{N}_{\rm ion}$ with the evolution of the UV luminosity density integrated down to $M_{\rm UV}=-13$ and concluded that in order for galaxies to reionize the IGM with a clumping fraction of $C_{\rm HII}=3$, a $\log(f_{\rm esc} \xi_{\rm ion}) = 24.50\pm 0.1$ is required. 
Assuming that our $\beta=-2.1$ subsample at $z=2$ is representative of galaxies at $z>6$, owing to their comparable blue UV slopes \citep{dunlop12,bouwens14}, we can use the \citet{bouwens15b} equation combined with our inferred {\xiion} value at $\beta = -2.1$ to predict an escape fraction.
The measured {\xiion} without any prior assumption for the $f_{\rm esc}$ (i.e., without assuming $E(B-V)$-derived $f_{\rm esc}$ that is described in Section~\ref{sec:xi_meas}), assuming that the Calzetti (SMC) curve for the UV dust correction is $\log(\xi_{\rm{ion}}/[{\rm{s^{-1}/erg\,s^{-1}\,Hz^{-1}}}])=25.32$ (25.37), results in an escape fraction of $0.15~(0.13)\pm 0.03$.
Our estimated escape fraction is more consistent with the low LyC escape fraction observed in $z\sim 2-4$ galaxies \citep[e.g.,][]{chen07,nestor13,siana15,grazian17,japelj17} than previously estimated escape fractions. 

UV-faint galaxies are thought to have a crucial contribution to cosmic reionization (due to their higher number density and possibly higher LyC escape fraction relative to the UV-bright galaxies; \citealt{ciardi12,duncan15,dijkstra16,anderson17,grazian17,japeli17}). An important question is whether the UV-faint galaxies also have higher {\xiion} values, which would be in favor of the significance of their role in reionization. 
We examine the relation between {\xiion} and UV absolute magnitude in Figure~\ref{fig:muv} for both the Calzetti and SMC curves with points from the higher-redshift study of \citet{bouwens16b} included. As seen in the figure, we do not find any significant trend between {\xiion} and UV luminosity. {\xiion} remains constant for galaxies below and above the characteristic UV magnitude at $1.9\leq z<2.7$, $M^*_{\rm AB}=-20.70$ \citep{reddy09}.

\begin{figure}[tbp]
	\centering
		\includegraphics[width=.49\textwidth,trim={.2cm 0 0 0},clip]{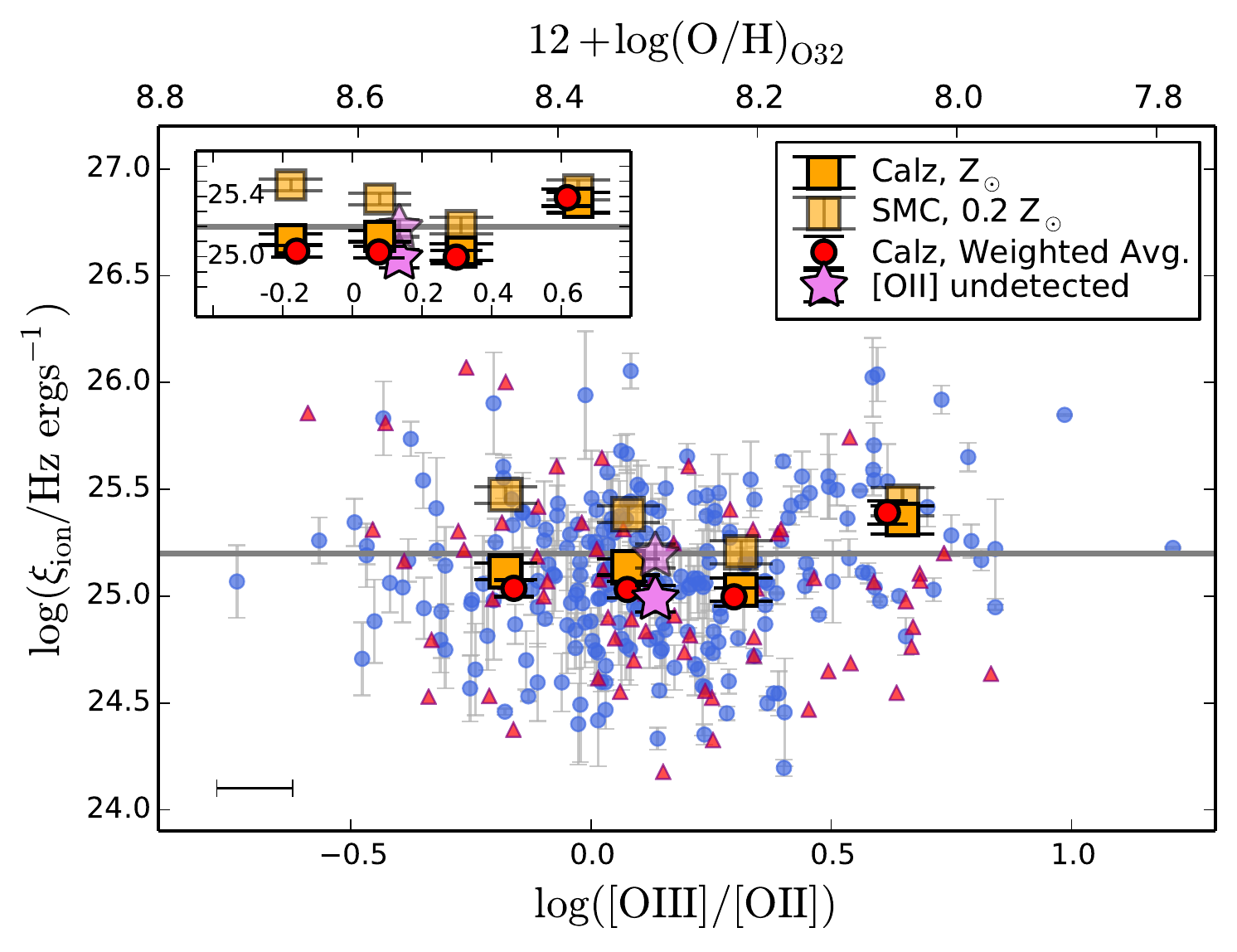}
		\caption{Dependence of {\xiion} on the [O{\sc iii}]/[O{\sc ii}] ratio.
		The orange and blue symbols are the same as in Figure~\ref{fig:beta} for the sample of {\halpha}, [O{\sc iii}], and [O{\sc ii}] detected galaxies.
		The red circles show the weighted averages of only the {\halpha}-{\hbeta} detected sample (the blue individual points), which are consistent with the orange stacks that include {\hbeta}-undetected galaxies as well.
		The violet star is the stack of undetected [O{\sc ii}] objects ([O{\sc iii}] detected), which does not display a higher {\xiion} compared to the rest of the stacks.
		The typical uncertainty in [O{\sc iii}]/[O{\sc ii}] is shown in the bottom-left corner.
		The {\o32} metallicity, based on the calibrations of \citealt{jones15}, is also displayed. 
		The horizontal grey line shows $\log(\xi_{\rm{ion}}/[{\rm{s^{-1}/erg\,s^{-1}\,Hz^{-1}}}])=25.2$ from \citet{robertson13}.
		}
		\label{fig:o32}
\end{figure}

\subsection{{\xiion} and ISM properties: {\o32} and the BPT Diagram}
\label{sec:gas}

The ionizing photon production efficiency, by definition, depends on the hardness and intensity of the source radiation field. MOSDEF includes measurements of multiple rest-optical emission lines whose ratios provide an independent way of constraining the properties of the ionizing radiation.
Measurements of these line ratios in galaxy-integrated MOSDEF spectra represent surface-brightness-weighted averages of the properties of star-forming regions.
In this section, we investigate the change in {\xiion} as a function of the [O{\sc iii}]/[O{\sc ii}] ratio ({\o32}), [O{\sc iii}]/{\hbeta} ratio (O3), and [N{\sc ii}]/{\halpha} ratio (N2), and across the [O{\sc iii}]/{\hbeta} versus [N{\sc ii}]/{\halpha} diagram \citep[BPT diagram;][]{bpt81}. 

As [O{\sc iii}] and [O{\sc ii}] are from the same element but with different ionization potentials, {\o32} is used as an empirical proxy for the ionization parameter in H{\sc ii} regions
 \citep{penston90,kewley02}. In Figure~\ref{fig:o32}, we show the relation between {\xiion} and {\o32}. In the Calzetti stacks, {\xiion} is constant at {\o32} $\lesssim 3$ and increases by $\sim 0.3$\,dex at {\o32} $\sim 5$. This increase is significant at the $> 3\,\sigma$ level.
The elevated {\xiion} at high {\o32} can be explained by high ionization parameters and low metallicities ({\o32} and oxygen abundances anticorrelate with each other; \citealt{jones15}). In such systems, the ionizing radiation is harder due to less metal blanketing, assuming that the gas-phase metallicity follows the stellar metallicity.
These high-{\o32} systems also have low masses, as {\o32} is anticorrelated with $M_*$ \citep[shown for the MOSDEF sample in][]{sanders16a}.
However, the trend of {\xiion} with {\o32} is different in the SMC stacks compared to the Calzetti stacks. The {\xiion} assuming an SMC curve for UV dust correction increases in both the low- and high-{\o32} bins (Figure~\ref{fig:o32}). The discrepancy between the Calzetti and the SMC results is largest in the lowest {\o32} bin, as these galaxies are also dustier, and hence the effect of any uncertainty in their dust correction is enhanced.
The different trends of the Calzetti and the SMC stacks show the importance of the systematic uncertainties associated with UV dust corrections (Section~\ref{sec:dustcurve}).

A high {\o32} may also be indicative of a density-bounded H{\sc ii} region \citep[see Figure 11 in][]{nakajima14,trainor16}, where the star-forming cloud is completely ionized and its radius is determined by the gas distribution and not by ionization equilibrium. 
In a density-bounded nebula, the outer layer with lower ionization species that produce [O{\sc ii}] is reduced, but the higher-ionization [O{\sc iii}] zone that forms preferentially at the inner radii of the H{\sc ii} region is unaffected.
Consequently, the ratio of [O{\sc iii}]/[O{\sc ii}] in a density-bounded nebula is larger than that of an ionization-bounded nebula.
On the other hand, a density-bounded nebula would have a higher leakage of ionizing photons (high $f_{\rm esc}$) compared to an ionization-bounded nebula \citep[see][]{giammanco05,jaskot13,nakajima14}. In this case, a fraction of the ionizing photons emitted from the source escape the nebula before producing recombination lines. 
As a result, if the high {\o32} ratios are to some extent associated with density-bounded systems \citep[see Figure 11 in][]{nakajima14}, their intrinsic {\xiion} is even higher than that inferred from {\halpha}. However, this effect is more pronounced in systems with very high {\o32} $\gtrsim 10$.

To ensure that the sample is not biased by limiting it to the objects with [O{\sc ii}] detection, we examine whether the galaxies with undetected [O{\sc ii}] are intrinsically different from the [O{\sc ii}]-detected objects by stacking the spectra of undetected [O{\sc ii}] objects (detected in [O{\sc iii}]). The stack is shown in Figure~\ref{fig:o32} and indicates {\o32} and {\xiion} similar to the rest of the sample with {\o32} $\lesssim 3$. The [O{\sc ii}] line in 35\% of the undetected [O{\sc ii}] objects lies very close to a sky line. This fraction for the detected [O{\sc ii}] sample is 25\%. A high fraction of contamination by sky lines implies that the [O{\sc ii}] line is not necessarily intrinsically faint in [O{\sc ii}]-undetected objects. 

\begin{figure}[tbp]
	\subfigure{
	\centering
		\includegraphics[width=.49\textwidth,trim={.2cm 0 0 0},clip]{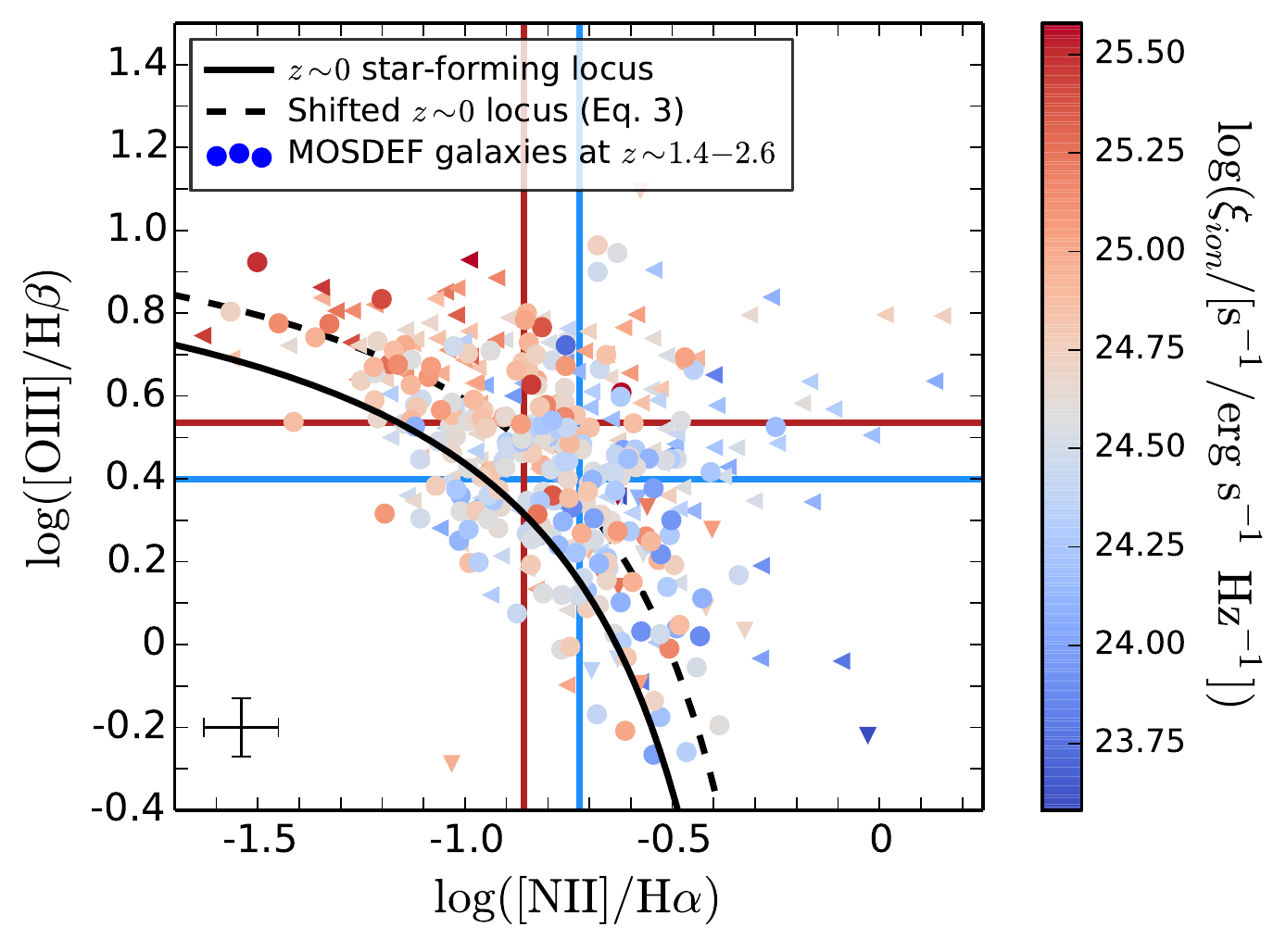}}
	\\
	\subfigure{
	\centering
		\includegraphics[width=.39\textwidth,trim={.2cm .2cm .2cm 0},clip]{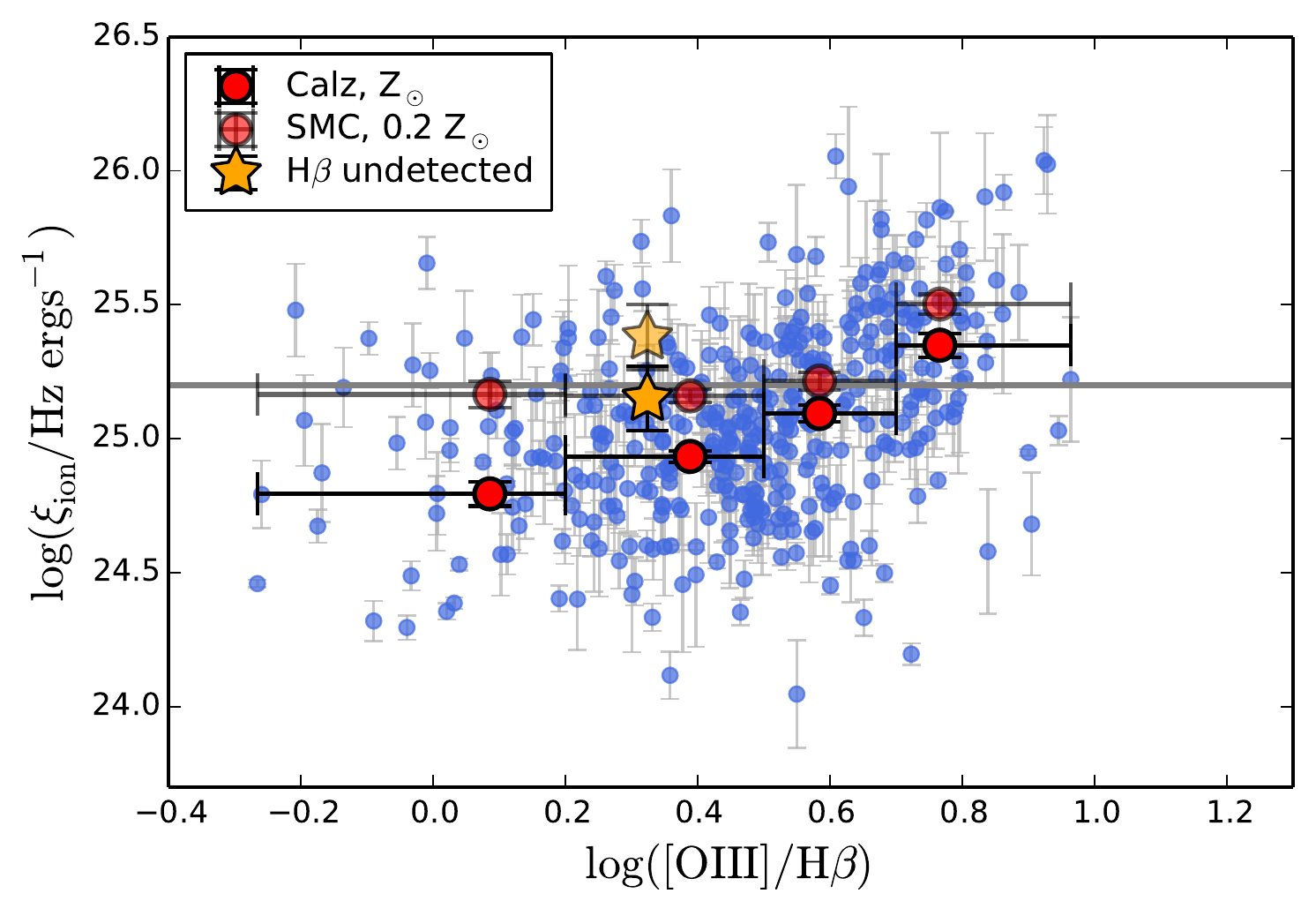}}
	\\
	\subfigure{
	\centering
		\includegraphics[width=.39\textwidth, trim={.2cm .2cm .2cm 0},clip]{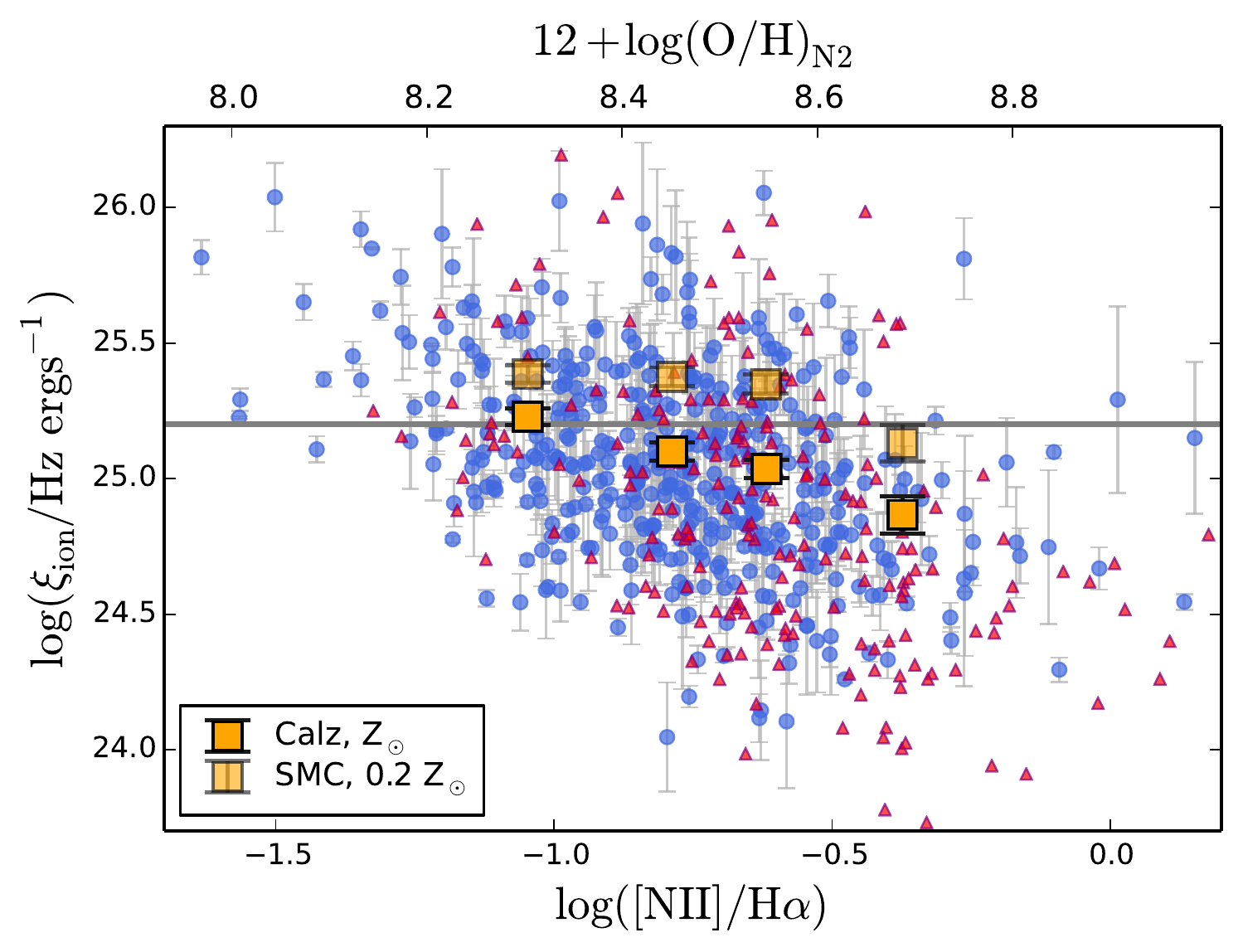}}	
		\caption{ ({\em Top:}) 
		Change in {\xiion} across the N2-BPT diagram. As a reference, we show the locus of star-forming galaxies at $z\sim 0$ \citep{kewley13} with a black solid line. 
		{\xiion} in is measured assuming of the Calzetti attenuation curve for UV dust correction.
		Circles denote objects with $> 3\,\sigma$ detections in all four lines. Downward triangles are $3\,\sigma$ upper limits on [O{\sc iii}]. Triangles pointing to the left show $3\,\sigma$ upper limits on [N{\sc ii}].
		The error bar in the corner indicates the typical uncertainties.
		{\xiion} is enhanced at high [O{\sc iii}]/{\hbeta} and low [N{\sc ii}]/{\halpha}. The horizontal and vertical lines show the average [O{\sc iii}]/{\hbeta} and [N{\sc ii}]/{\halpha}, respectively, for the upper (red) and lower (blue) half of the {\xiion} distribution. The distribution is the same for the sample with only $z>2$ galaxies. The $z<2$ sample is not large enough to statistically constrain the relation.
		Also shown is the shifted $z\sim 0$ sequence with a dashed curve, such that the $z\sim 2$ sample is equally distributed above and below the curve.
		({\em Middle:}) Dependence of {\xiion} on [O{\sc iii}]/{\hbeta}. The red symbols are weighted averages of {\xiion} in bins of [O{\sc iii}]/{\hbeta}. 
		The horizontal line shows $\log(\xi_{\rm{ion}}/[{\rm{s^{-1}/erg\,s^{-1}\,Hz^{-1}}}])=25.2$ from \citet{robertson13}.
		({\em Bottom:}) Dependence of {\xiion} on [N{\sc ii}]/{\halpha}. The orange symbols are measured from the composite spectra in bins of [N{\sc ii}]/{\halpha}. 
		The N2 metallicity, based on the calibrations of \citet{pp04}, is also shown.
		}
		\label{fig:bpt}
\end{figure}

We also consider the variations of {\xiion} with location in the [O{\sc iii}]/{\hbeta} versus [N{\sc ii}]/{\halpha} BPT diagram (hereafter the N2-BPT diagram). The top panel of Figure~\ref{fig:bpt} shows galaxies in our sample in the N2-BPT diagram, color coded by $\log(\xi_{\rm ion})$.
The position of galaxies in the N2-BPT diagram is indicative of both the stellar ionizing radiation field and the conditions of the gas in the star-forming regions. A higher ionization parameter increases O3 and keeps N2 almost constant at N2 $\lesssim -0.9$ \citep[low metallicity;][]{kewley13}. In agreement with this statement, we also see in the top panel of Figure~\ref{fig:bpt} that {\xiion} increases with increasing O3 at low N2.
On the other hand, a {\em harder} ionizing radiation field leads to higher [N{\sc ii}]/{\halpha} and [O{\sc iii}]/{\hbeta} at fixed nebular metallicity (i.e., a shift in the star-forming sequence in the BPT diagram toward larger [N{\sc ii}]/{\halpha} and [O{\sc iii}]/{\hbeta}). Some studies have suggested that the offset of $z\sim 2$ galaxies from the locus of $z\sim 0$ star-forming galaxies in the N2-BPT diagram is due to a harder stellar ionizing radiation field \citep{steidel14,steidel16,strom17}.
To examine whether {\xiion} varies with the offset from the BPT star-forming sequence, we split the sample with detected lines into two bins, according to the $z\sim 0$ locus (adopted from \citealt{kewley13}) shifted toward higher O3 and higher N2. The new shifted BPT sequence for our $z\sim 2$ sample is
\begin{equation}
\log(\frac{\rm [O{\normalfont\textsc{iii}}]}{\rm H\beta}) = \frac{0.61}{\log({\rm N{\normalfont\textsc{ii}}/H\alpha})+0.005} + 1.2.
\end{equation}
There are 125 galaxies on each side of the shifted sequence, shown with dashed curve in Figure~\ref{fig:bpt}. The average {\xiion} above and below the shifted sequence are, respectively, $\log(\xi_{\rm{ion}}/[{\rm{s^{-1}/erg\,s^{-1}\,Hz^{-1}}}])= 25.07\pm 0.03$ and $25.04\pm 0.03$, suggesting no change in the hardness of the ionizing radiation with the offset from the $z\sim 0$ sequence. The other possible cause of the offset between the locus of star-forming galaxies in the BPT diagram at $z\sim 2$ and $z\sim 0$ galaxies is a higher N$/$O at a given O$/$H in high-redshift galaxies compared to that of the local ones \citep[for detailed discussions see][]{shapley15,masters16,cowie16,sanders16a}.

We also show the trends of {\xiion} with O3 and N2 separately in the middle and bottom panels of Figure~\ref{fig:bpt}, where {\xiion} increases with increasing O3 and decreasing N2. {\xiion} varies by 0.36\,dex from $\sim 0.4-1.0\,Z_{\odot}$, based on the N2 metallicity indicator shown in the bottom panel of Figure~\ref{fig:bpt}. This change is significantly larger than that predicted from the SPS models (0.17\,dex from $\sim 0.07-1.5\,Z_{\odot}$; Section~\ref{sec:sps}), assuming that the stellar and gas-phase metallicities correlate with each other. 
The strong correlation that we observe between {\xiion} and the N2 metallicity is partly due to the fact that {\halpha} is used in both {\xiion} (in the numerator) and N2 (in the denominator).
We note that {\xiion} changes less significantly with the {\o32} metallicity ($\sim 0.3$\,dex; Figure~\ref{fig:o32}). The {\xiion} and {\o32} calculations are less dependent, as the only common term in {\xiion} and {\o32} is the nebular dust attenuation correction ({\halpha}/{\hbeta}). 

In summary, Figures~\ref{fig:o32} and \ref{fig:bpt} indicate an increase of the production efficiency of ionizing photons in star-forming galaxies with high ionization parameter (high O3, Figure~\ref{fig:bpt}; high {\o32}, Figure~\ref{fig:o32}) and at low metallicities (low N2, Figure~\ref{fig:bpt}). 
The result is further shown with the shift in the mean O3 and N2 for the upper and lower halves of the {\xiion} distribution (the solid vertical and horizontal lines in Figure~\ref{fig:bpt}), which is consistent with a shift toward higher metallicity (lower O3, higher N2) at lower {\xiion}.  
There are other interrelated parameters that may play a role in the observed {\xiion} trend with metallicity, such as mass and star-formation history. Galaxies with higher metallicities tend to have higher masses \citep[see the MOSDEF mass-metallicity study;][]{sanders16a} and a higher fraction of older and lower-mass stars, which lead to a lower {\xiion}.

\begin{figure}[tbp]
	\centering
		\includegraphics[width=.49\textwidth,trim={.2cm 0 0 0},clip]{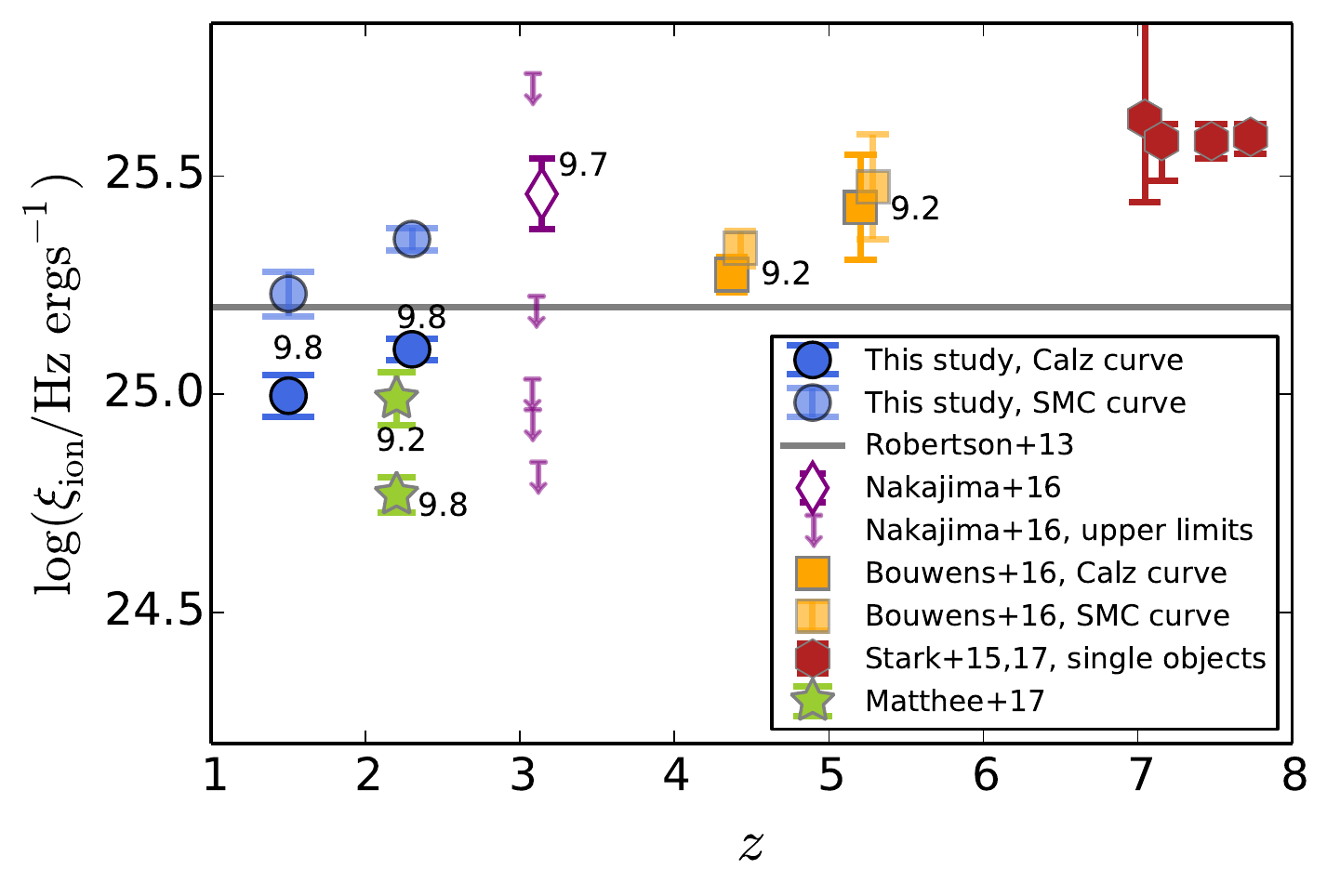}
		\caption{Ionizing photon production efficiency measured over a wide range of redshifts. Our measurements in two redshift bins are shown with blue symbols. We include the inferred {\xiion} values from the literature at different redshifts. For the details of each study, refer to the text.
		The numbers next to the markers indicate the average masses of the samples. We limit our MOSDEF samples to $\log(M_*/{\rm M_{\odot}})<10.3$ to obtain a mass distribution more similar to that of other high-redshift studies.
		}
		\label{fig:redshift}
\end{figure}

\section{Comparison with Other Studies}
\label{sec:z}

One motivation for constraining {\xiion} is that it is an important input to cosmic reionization models. To see whether our estimates of {\xiion} at $z\sim 2$ can be extended to $z\gtrsim 6$ galaxies, we compare our results with the studies of higher redshift galaxies from the literature in Figure~\ref{fig:redshift}. 

To make our sample more comparable to that of the higher redshift studies, we only use galaxies with $\log(M_*/{\rm M_{\odot}})<10.3$, which results in 141 and 315 galaxies at $z\leq 2.0$ and $z>2.0$, respectively. The upper-mass limit is chosen to match the maximum mass limit of the \citet{bouwens16b} sample at $z\sim 4.4$. Our limited sample still has a higher average mass than that of the \citet{bouwens16b} sample, as we lack galaxies with $\log(M_*/{\rm M_{\odot}})\lesssim 9.0$. The numbers next to the markers in Figure~\ref{fig:redshift} indicate the average masses of each sample. As is apparent, each sample traces galaxies over different stellar mass ranges.

\begin{figure}[tbp]
	\centering
		\includegraphics[width=.49\textwidth,trim={.2cm 0 0 0},clip]{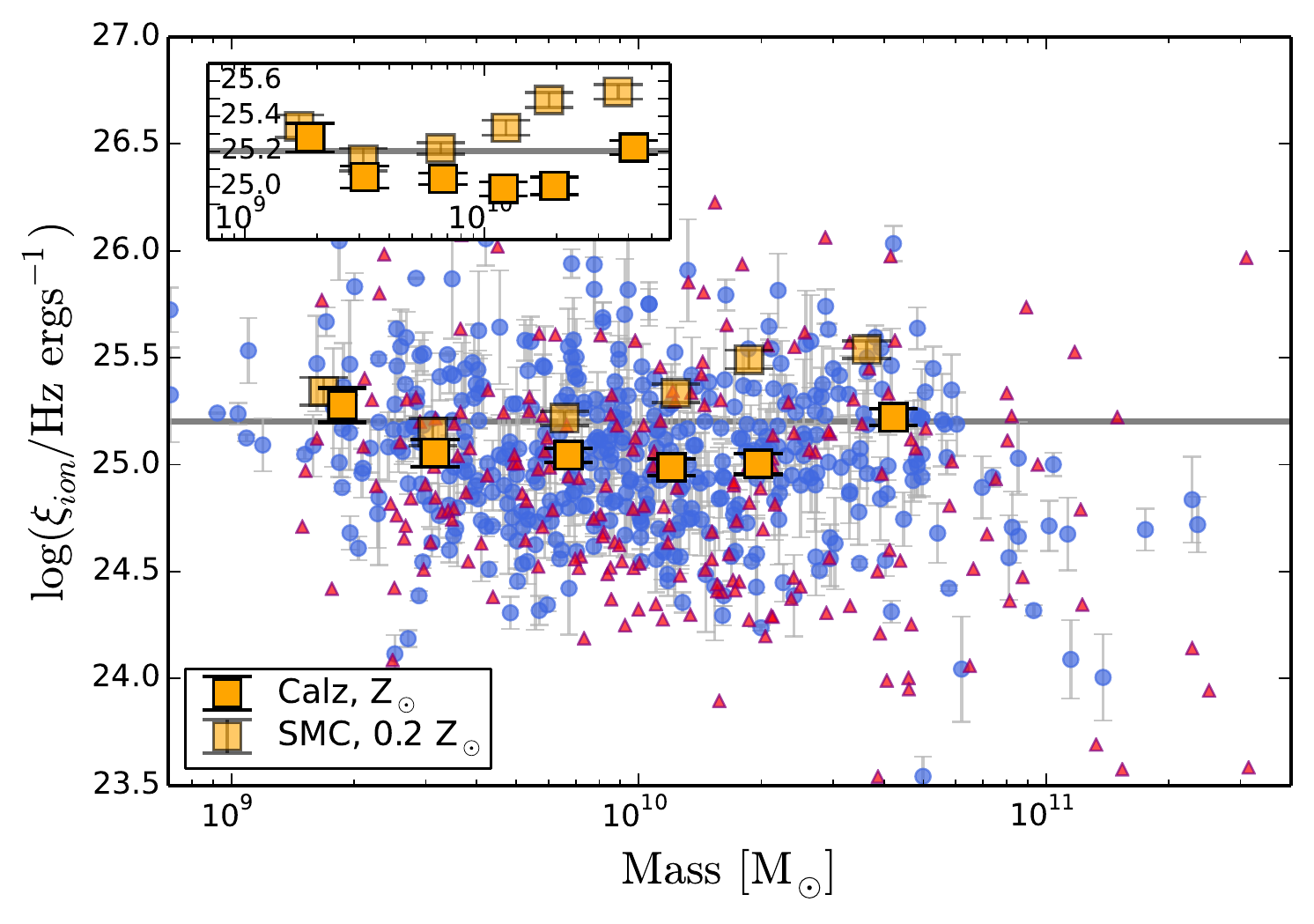}
		\caption{Dependence of {\xiion} on stellar mass.
		The symbols are the same as in Figure~\ref{fig:beta}.
		}
		\label{fig:mass}
\end{figure}

To determine to what extent {\xiion} is affected by stellar mass, we evaluate {\xiion} in bins of mass in Figure~\ref{fig:mass}. {\xiion} is almost constant at high masses but the $0.23\pm 0.10$\,dex increase in the lowest mass bin may be indicative of a mass dependency of {\xiion}. 
The weak trend with stellar mass is also seen in predictions from simulations \citep{wilkins16}.
We cannot test whether the observed increase in {\xiion} at lower masses persists, as the MOSDEF sample is only complete down to $10^{9.5}$\,\msun \citep{shivaei15b}.
However, the systematic differences among the samples in Figure~\ref{fig:redshift} are not limited to their mass distributions. There are other parameters interrelated with mass and with each other, such as metallicity, specific SFR, and age, that potentially affect the trend we observe in Figure~\ref{fig:redshift}. We review the details of the studies presented in Figure~\ref{fig:redshift} and their potential differences in the following paragraphs.

In Figure~\ref{fig:redshift}, we also show the results of \citet{matthee17} at $z\sim 2$, which are based on a large sample of {\halpha} emitters from a narrowband survey to estimate {\xiion}. The two {\xiion} values are calculated for two subsamples with different stellar masses, indicated in black font next to the markers.
\citet{matthee17} did not have {\hbeta} measurements, and hence the authors applied the same attenuation toward the stellar and nebular light, derived from a locally calibrated relation between the stellar dust attenuation and mass \citep{garnbest10}. A higher nebular reddening compared to the stellar reddening would increase their {\xiion} values.
As shown in both \citet{matthee17} and this work, the average {\xiion} is highly affected by the dust correction method, which is probably the main source of the discrepancy seen between the results of \citet{matthee17} and ours at $z\sim 2$.

\citet{nakajima16} used {\hbeta} luminosity with an assumption of no dust attenuation to estimate {\xiion} for a small sample of 15 Ly$\alpha$ emitters (LAEs) and Lyman break galaxies (LBGs) at $z\sim 3$. Out of 15 objects, 12 were LAEs (plus one AGN and two LBGs), which poses a strong sample selection bias, as LAEs occupy a different part of the BPT diagram and have different ionization properties \citep[e.g.,][]{nakajima13,erb16}.
We show the average {\xiion} of the detected {\hbeta} sample along with the $3\,\sigma$ upper limits of the undetected ones in Figure~\ref{fig:redshift}. These authors discussed that the assumption of the same stellar and nebular color-excess, the median $E(B-V)=0.03$ derived from SED modeling with an SMC curve, decreases their ${\xiion}$ values by $\sim 0.1$\,dex. 

At higher redshifts, \citet{bouwens16b} estimated {\halpha} from IRAC photometry and calculated {\xiion} at $z\sim 4-5$. In the absence of {\hbeta} measurements, {\halpha} luminosities were dust-corrected using stellar reddening derived from SED fitting. As the UV and {\halpha} luminosities were dust corrected assuming the same $E(B-V)$ values, the difference between the Calzetti and SMC values of \citet{bouwens16b} is smaller than that of our measurements.

Also shown in Figure~\ref{fig:redshift} are the {\xiion} inferred from UV metal lines for four individual luminous,
young, and metal-poor galaxies at $z\sim 7$ from \citet{stark15} and \citet{stark17}.
The elevated {\xiion} values are not necessarily typical of high-redshift galaxies, as the galaxies in the \citet{stark15} and \citet{stark17} studies were selected to have extremely large [O{\sc iii}]$+${\hbeta} equivalent widths. The galaxies also show Ly$\alpha$ detections, although the IGM is expected to be significantly neutral at $z=7-9$. \citet{stark17} speculated that the transmission of Ly$\alpha$ in these objects is likely due to their hard radiation field, which has reduced the covering fraction of neutral hydrogen in the circumgalactic medium. The unique selection criteria of these galaxies make it nontrivial to compare them with our sample at $z\sim 2$ in the context of {\xiion} redshift evolution.

Future studies with {\em JWST} will be particularly useful to extend the dynamic range of studies of {\xiion} to galaxies with lower masses, lower metallicities, and younger ages, and to constrain the redshift evolution of {\xiion} in a more robust and consistent way. However, as discussed in Section~\ref{sec:dustcurve}, the UV dust attenuation curve remains a significant uncertainty in assessing {\xiion}.

\section{Summary}
\label{sec:summary}

We use a sample of 673 galaxies from the MOSDEF survey with {\halpha} and {\hbeta} spectra and UV continuum photometry to constrain the hydrogen-ionizing photon production efficiency, {\xiion}, at $z= 1.37-2.61$. 
We convert dust-corrected {\halpha} to the production rate of ionizing photons. We then divide this quantity by the UV luminosity at 1500\,\AA, dust-corrected using the SED-derived reddening, to estimate {\xiion}.
The main results are as follows.

\begin{itemize}

\item We find an average {\xiion} value of $\log(\xi_{\rm{ion}}/[{\rm{s^{-1}/erg\,s^{-1}\,Hz^{-1}}}])=25.10~(25.34)$ for our entire sample at $z=1.37-2.61$, assuming the Calzetti (SMC) curve for the UV dust correction (Section~\ref{sec:scatter}). After measurement uncertainties are taken into account, the {\xiion} distribution has a large scatter of 0.28\,dex. 
Dividing the sample into two bins of redshift, the average {\xiion} at $z\sim 1.5$ and $z\sim 2.3$ are 25.05 and 25.13, respectively, assuming the Calzetti curve (an SMC curve results in values $\sim 0.24$\,dex larger). These results imply that {\xiion} does not evolve with redshift (from $z\sim 1.5$ to 2.3) significantly.
The average {\xiion} values that we derive assuming the Calzetti and SMC curves bracket the canonical {\xiion} of 25.20  \citep{robertson13} that is commonly used in the reionization models.

\item We investigate the sources of the large {\em intrinsic} (i.e., measurement uncertainty subtracted) scatter of 0.28\,dex in the {\xiion} distribution by investigating the galaxy-to-galaxy variations in the dust attenuation curve, IMF, stellar metallicity, star formation burstiness, and binary/single evolutionary paths. We conclude that the galaxy-to-galaxy variations in the dust attenuation curve introduce a $\sim 0.1-0.2$\,dex uncertainty in the scatter (Section~\ref{sec:dustcurve}).
We further investigate the effect of the galaxy-to-galaxy variations in the stellar population properties of our galaxies on the scatter (Section~\ref{sec:sps}), as follows.
Binary evolution increases {\xiion} predictions from SPS models by 0.17\,dex.
Variations in the upper-mass cutoff and the slope of the IMF cause a large variation in {\xiion} of 0.58\,dex, as a higher IMF upper-mass cutoff and shallower slope increases the fraction of hot and massive ionizing stars.
{\xiion} also varies with stellar metallicity (by 0.17\,dex from $Z=0.07$ to 1.5\,$Z_{\odot}$) and is higher in metal-poor galaxies, mainly due to the decreased effect of metal blanketing. 
A starburst with an amplitude of five times the underlying SFR can change {\xiion} from $\log(\xi_{\rm{ion}}/[{\rm{s^{-1}/erg\,s^{-1}\,Hz^{-1}}}])=24.7$ to 25.3. However, a high frequency of strong bursts is unlikely for most of our galaxies with masses $\gtrsim 10^{9.5}$\,\msun.

\item We demonstrate that even with robust dust-corrected {\halpha} measurements and $N({\rm H}^0)$ estimates, the inferred {\xiion} values suffer from large systematic uncertainties that originate from the uncertainties in the assumed UV dust attenuation curve (Section~\ref{sec:dustcurve}). An SMC curve with an intrinsically blue UV slope (a result of the subsolar metallicity assumption for the stellar population) systematically increases the {\xiion} estimates by as large as 0.4\,dex compared to the results of the shallower Calzetti curve with solar abundances. The systematic change is shown in all Figures~\ref{fig:beta}$-$\ref{fig:mass}. A more realistic mass-dependent UV dust attenuation curve, where the steepness of the curve anticorrelates with the stellar mass, would affect our results and enhance the {\xiion} variations we observe as a function of mass and $\beta$. This systematic uncertainty is an important factor to consider in future rest-optical observations with {\em JWST}.

\item We find an enhanced value of {\xiion} in galaxies with a blue observed UV slope ($\log(\xi_{\rm{ion}}/[{\rm{s^{-1}/erg\,s^{-1}\,Hz^{-1}}}])=25.37\pm 0.08$ at $\langle \beta\rangle=-2.1$, Section~\ref{sec:uv}). This value is calculated using an SMC curve, and without any prior correction for the LyC escape fraction.
If galaxies at $z>6$ are similar to the ones in our sample with blue $\beta$ and low masses (an assumption based on their similar UV slopes), the higher {\xiion} of 25.37 relative to the canonical {\xiion} of 25.20 suggests that a {\em lower} LyC escape fraction is required in order for galaxies to keep the universe ionized at such early times.

\item We show that {\xiion} is almost constant at {\o32} $\lesssim 3$ and increases by $\sim 0.2-0.3$\,dex at {\o32} $\sim 5$ to $\log(\xi_{\rm{ion}}/[{\rm{s^{-1}/erg\,s^{-1}\,Hz^{-1}}}])=25.36\pm 0.07$ (Figure~\ref{fig:o32}).
We also find a strong anticorrelation between {\xiion} and N2 (Figure~\ref{fig:bpt}). These two observations indicate an increase of {\xiion} in low-metallicity environments.
Furthermore, we find that in the N2-BPT diagram, {\xiion} does not change with the offset from the $z\sim 0$ star-forming locus, suggesting no change in the hardness of the ionizing radiation with the offset from the $z\sim 0$ sequence (Section~\ref{sec:gas}).

\item We do not find a strong correlation between {\xiion} and stellar mass. However, there is a $0.23\pm 0.10$\,dex increase of {\xiion} in the lowest mass bin with $\langle M_*\rangle=10^{9.3}$\,\msun (Figure~\ref{fig:mass}). Samples with a wider range in mass that span to $M_*<10^9$\,\msun~are needed to complete the picture.

\item We compare our results at $z\sim 1.5$ and 2.3 with other high-redshift studies in the literature (Section~\ref{sec:z}). Samples at higher redshifts are characterized by higher average {\xiion}. As the samples have different stellar masses, metallicities, ages, and dust distributions, it is not trivial to conclude robustly whether the increase of {\xiion} with redshift in Figure~\ref{fig:redshift} is indicative of {\xiion} redshift evolution or is caused by sample selection biases.

\end{itemize}

Although, as mentioned, the uncertainties in the UV dust correction make it difficult to robustly infer {\xiion}, the uncertainties are significantly reduced for galaxies with low dust content (e.g., see the two Calzetti and SMC points with the bluest $\beta$ in Figure~\ref{fig:beta}). These less dusty galaxies also have the highest ionizing photon production efficiencies and are possibly similar to the faint galaxies at $z>6$ that have a crucial contribution to cosmic reionization.
Therefore, future observations with the {\em JWST} NIRSpec and NIRCam instruments to probe higher redshift and lower-mass galaxies would be highly valuable to better constrain the evolution of {\xiion}.
Moreover, to diminish the UV dust attenuation uncertainties, the studies of rest-frame UV spectra (e.g., modeling the observed P Cygni absorption features that are sensitive to the presence of massive stars) will be an independent way of inferring {\xiion} and exploring the physical sources of the variations in {\xiion} from galaxy to galaxy.

\vspace{5mm}
I.S. thanks Rychard Bouwens and George Rieke for helpful comments.
Funding for the MOSDEF survey is provided by NSF AAG grants AST-1312780, 1312547, 1312764, and 1313171 and archival grant AR-13907, provided by NASA through a grant from the Space Telescope Science Institute.
The data presented herein were obtained at the W.M. Keck Observatory, which is operated as a scientific partnership among the California Institute of Technology, the University of California, and the National Aeronautics and Space Administration. The Observatory was made possible by the generous financial support of the W.M. Keck Foundation.
The authors wish to recognize and acknowledge the very significant cultural role and reverence that the summit of Maunakea has always had within the indigenous Hawaiian community. We are most fortunate to have the opportunity to conduct observations from this mountain.


\end{document}